\documentclass[11pt]{article}
\usepackage{times}
\usepackage[paperwidth=199.8mm,
paperheight=297mm,centering,hmargin=25mm,vmargin=25mm]{geometry}
\usepackage{authblk} 
\usepackage[bottom]{footmisc} 

\usepackage[titletoc,title]{appendix} 
\usepackage{changepage}

\usepackage{hyperref}
\hypersetup{colorlinks=true,citecolor=myblue,linkcolor=myblue,
filecolor=myblue,urlcolor=myblue,breaklinks=true}
\usepackage{url}
\usepackage[dvipsnames]{xcolor}
\usepackage{color}
\usepackage{framed}

\definecolor{shadecolor}{rgb}{0.9,0.9,0.9}
\definecolor{mylightgray}{RGB}{100,100,100}

\definecolor{myblue}{RGB}{0, 68, 116}
\definecolor{mycyan}{RGB}{0, 97, 91}
\definecolor{mygreen}{RGB}{2, 102, 1}
\definecolor{myorange}{RGB}{240, 102, 0}
\definecolor{myred}{RGB}{172, 23, 0}
\definecolor{mymagenta}{RGB}{140,16,73}

\usepackage{mathtools}
\usepackage{amsmath}
\usepackage{amssymb}
\usepackage{bm}
\usepackage[shortlabels]{enumitem}
\usepackage{graphicx,epic,eepic,epsfig,latexsym,verbatim}
\usepackage{dsfont}
\usepackage{mathrsfs}

\usepackage{subcaption}
\usepackage{caption}
\usepackage{float}
\usepackage{placeins}
\usepackage{tikz}
\usetikzlibrary{arrows.meta}
\usepackage{relsize}
\usepackage{pgfplots}
\pgfplotsset{compat=1.18}

\usepackage{amsthm}
\usepackage{tcolorbox}
\tcbuselibrary{skins}
\tcbuselibrary{breakable}

\makeatletter
\newcommand{\DefineColoredTheoremStyle}[2]{%
  \newtheoremstyle{#1}
    {3pt}{3pt}
    {\itshape}
    {}
    {\bfseries\color{#2}}
    {\;}
    { }
    {\thmname{##1}\thmnumber{ ##2}\textnormal{\thmnote{ (##3)}}}
}
\makeatother

\DefineColoredTheoremStyle{lemmstyle}{black}
\DefineColoredTheoremStyle{propstyle}{black}
\DefineColoredTheoremStyle{thmstyle}{black}
\DefineColoredTheoremStyle{corostyle}{black}
\DefineColoredTheoremStyle{conjstyle}{black}
\DefineColoredTheoremStyle{problstyle}{black}
\DefineColoredTheoremStyle{plainstyle}{black}

\theoremstyle{lemmstyle}\newtheorem{lemma}{Lemma}[section]
\theoremstyle{propstyle}\newtheorem{prop}{Proposition}[section]
\theoremstyle{thmstyle}\newtheorem{theorem}{Theorem}[section]
\theoremstyle{corostyle}\newtheorem{corollary}{Corollary}[section]
\theoremstyle{conjstyle}\newtheorem{conjecture}{Conjecture}[section]
\theoremstyle{problstyle}\newtheorem{problem}{Problem}[section]
\theoremstyle{plainstyle}\newtheorem{definition}{Definition}[section]
\theoremstyle{plainstyle}\newtheorem{assumption}{Assumption}[section]
\theoremstyle{plainstyle}\newtheorem{example}{Example}[section]
\theoremstyle{plainstyle}\newtheorem{remark}{Remark}[section]

\tcbset{
    commonstyle/.style n args={1}{
        oversize=0.5mm,
        colback={#1!8},
        colframe={#1!10},
        boxrule=1pt,
        boxsep=2mm,
        left=1mm,
        right=1mm,
        top=2mm,
        bottom=2mm,
        breakable
    }
}

\newenvironment{boxlemma}{%
  \begin{tcolorbox}[commonstyle={mycyan}]%
    \begin{lemma}%
}{%
    \end{lemma}%
  \end{tcolorbox}%
}

\newenvironment{boxproposition}{%
  \begin{tcolorbox}[commonstyle={myblue}]%
    \begin{prop}%
}{%
    \end{prop}%
  \end{tcolorbox}%
}

\newenvironment{boxtheorem}{%
  \begin{tcolorbox}[commonstyle={mymagenta}]%
    \begin{theorem}%
}{%
    \end{theorem}%
  \end{tcolorbox}%
}

\newenvironment{boxcorollary}{%
  \begin{tcolorbox}[commonstyle={myorange}]%
    \begin{corollary}%
}{%
    \end{corollary}%
  \end{tcolorbox}%
}

\usepackage{colortbl}
\usepackage{pifont} 
\usepackage{booktabs}
\usepackage{makecell} 
\usepackage{multirow}
\usepackage{longtable}
\usepackage{array}
\usepackage{ragged2e}
\newcolumntype{P}[1]{>{\RaggedRight\arraybackslash}p{#1}}
\newcommand{\nc}{\newcommand}
\nc{\rnc}{\renewcommand}

\nc{\<}{\langle}
\rnc{\>}{\rangle}
\nc{\bra}[1]{\langle#1|}
\nc{\ket}[1]{|#1\rangle}
\nc{\ketbra}[2]{|#1\rangle\!\langle#2|}
\nc{\braket}[2]{\langle#1|#2\rangle}
\nc{\braandket}[3]{\langle #1|#2|#3\rangle}
\nc{\proj}[1]{| #1\rangle\!\langle #1 |}
\nc{\avg}[1]{\langle#1\rangle}

\nc{\rank}{\operatorname{Rank}}
\nc{\id}{{\operatorname{id}}}
\nc{\iid}{{\operatorname{iid}}}
\nc{\supp}{{\operatorname{supp}}}
\nc{\smfrac}[2]{\mbox{$\frac{#1}{#2}$}}
\nc{\tr}{\operatorname{Tr}}
\nc{\ox}{\otimes}
\nc{\floor}[1]{\lfloor #1 \rfloor}
\nc{\trans}{\mathsf T}
\nc{\img}{\mathbf{i}}

\makeatletter

\newcommand{\solidgamearrowpicture@}[3]{%
    \tikz[x=1ex,y=1ex,line cap=round,line join=round]
      \draw[line width=#3,-{Triangle[length=0.64ex,width=0.68ex]}] (0,0) -- (#1,#2);%
}
\newcommand{\opengamearrowpicture@}[3]{%
    \tikz[x=1ex,y=1ex,line cap=round,line join=round]
      \draw[line width=#3,-{Triangle[open,length=0.64ex,width=0.68ex]}] (0,0) -- (#1,#2);%
}
\newsavebox{\solidgameuparrow@display}
\newsavebox{\solidgameuparrow@text}
\newsavebox{\solidgameuparrow@script}
\newsavebox{\solidgameuparrow@scriptscript}
\newsavebox{\solidgamedownarrow@display}
\newsavebox{\solidgamedownarrow@text}
\newsavebox{\solidgamedownarrow@script}
\newsavebox{\solidgamedownarrow@scriptscript}
\newsavebox{\opengameuparrow@display}
\newsavebox{\opengameuparrow@text}
\newsavebox{\opengameuparrow@script}
\newsavebox{\opengameuparrow@scriptscript}
\newsavebox{\opengamedownarrow@display}
\newsavebox{\opengamedownarrow@text}
\newsavebox{\opengamedownarrow@script}
\newsavebox{\opengamedownarrow@scriptscript}
\AtBeginDocument{%
  \sbox{\solidgameuparrow@display}{\solidgamearrowpicture@{0}{1.62}{0.58pt}}%
  \sbox{\solidgameuparrow@text}{\solidgamearrowpicture@{0}{1.50}{0.52pt}}%
  \sbox{\solidgameuparrow@script}{\solidgamearrowpicture@{0}{1.30}{0.44pt}}%
  \sbox{\solidgameuparrow@scriptscript}{\solidgamearrowpicture@{0}{1.12}{0.38pt}}%
  \sbox{\solidgamedownarrow@display}{\solidgamearrowpicture@{0}{-1.62}{0.58pt}}%
  \sbox{\solidgamedownarrow@text}{\solidgamearrowpicture@{0}{-1.50}{0.52pt}}%
  \sbox{\solidgamedownarrow@script}{\solidgamearrowpicture@{0}{-1.30}{0.44pt}}%
  \sbox{\solidgamedownarrow@scriptscript}{\solidgamearrowpicture@{0}{-1.12}{0.38pt}}%
  \sbox{\opengameuparrow@display}{\opengamearrowpicture@{0}{1.62}{0.58pt}}%
  \sbox{\opengameuparrow@text}{\opengamearrowpicture@{0}{1.50}{0.52pt}}%
  \sbox{\opengameuparrow@script}{\opengamearrowpicture@{0}{1.30}{0.44pt}}%
  \sbox{\opengameuparrow@scriptscript}{\opengamearrowpicture@{0}{1.12}{0.38pt}}%
  \sbox{\opengamedownarrow@display}{\opengamearrowpicture@{0}{-1.62}{0.58pt}}%
  \sbox{\opengamedownarrow@text}{\opengamearrowpicture@{0}{-1.50}{0.52pt}}%
  \sbox{\opengamedownarrow@script}{\opengamearrowpicture@{0}{-1.30}{0.44pt}}%
  \sbox{\opengamedownarrow@scriptscript}{\opengamearrowpicture@{0}{-1.12}{0.38pt}}%
}
\newcommand{\gamearrowbox@}[1]{%
  \mathrel{\vcenter{\hbox{%
    \usebox{#1}%
  }}}%
}
\DeclareRobustCommand{\soliduparrow}{%
  \mathchoice
    {\gamearrowbox@{\solidgameuparrow@display}}%
    {\gamearrowbox@{\solidgameuparrow@text}}%
    {\gamearrowbox@{\solidgameuparrow@script}}%
    {\gamearrowbox@{\solidgameuparrow@scriptscript}}%
}
\DeclareRobustCommand{\soliddownarrow}{%
  \mathchoice
    {\gamearrowbox@{\solidgamedownarrow@display}}%
    {\gamearrowbox@{\solidgamedownarrow@text}}%
    {\gamearrowbox@{\solidgamedownarrow@script}}%
    {\gamearrowbox@{\solidgamedownarrow@scriptscript}}%
}
\DeclareRobustCommand{\emptyuparrow}{%
  \mathchoice
    {\gamearrowbox@{\opengameuparrow@display}}%
    {\gamearrowbox@{\opengameuparrow@text}}%
    {\gamearrowbox@{\opengameuparrow@script}}%
    {\gamearrowbox@{\opengameuparrow@scriptscript}}%
}
\DeclareRobustCommand{\emptydownarrow}{%
  \mathchoice
    {\gamearrowbox@{\opengamedownarrow@display}}%
    {\gamearrowbox@{\opengamedownarrow@text}}%
    {\gamearrowbox@{\opengamedownarrow@script}}%
    {\gamearrowbox@{\opengamedownarrow@scriptscript}}%
}
\protected\def\uparrow{\soliduparrow}
\protected\def\downarrow{\soliddownarrow}
\makeatother

\nc{\cA}{{\cal A}}
\nc{\cB}{{\cal B}}
\nc{\cC}{{\cal C}}
\nc{\cD}{{\cal D}}
\nc{\cE}{{\cal E}}
\nc{\cF}{{\cal F}}
\nc{\cG}{{\cal G}}
\nc{\cH}{{\cal H}}
\nc{\cI}{{\cal I}}
\nc{\cJ}{{\cal J}}
\nc{\cK}{{\cal K}}
\nc{\cL}{{\cal L}}
\nc{\cM}{{\cal M}}
\nc{\cN}{{\cal N}}
\nc{\cO}{{\cal O}}
\nc{\cP}{{\cal P}}
\nc{\cQ}{{\cal Q}}
\nc{\cR}{{\cal R}}
\nc{\cS}{{\cal S}}
\nc{\cT}{{\cal T}}
\nc{\cV}{{\cal V}}
\nc{\cU}{{\cal U}}
\nc{\cX}{{\cal X}}
\nc{\cY}{{\cal Y}}
\nc{\cZ}{{\cal Z}}
\nc{\cW}{{\cal W}}

\nc{\bp}{\boldsymbol{p}}
\nc{\bq}{\boldsymbol{q}}
\nc{\brho}{\boldsymbol{\rho}}
\nc{\bsigma}{\boldsymbol{\sigma}}
\nc{\bomega}{\boldsymbol{\omega}}
\nc{\bmu}{\boldsymbol{\mu}}
\nc{\bT}{\boldsymbol{T}}
\nc{\bS}{\boldsymbol{S}}

\nc{\RR}{{{\mathbb R}}}
\nc{\CC}{{{\mathbb C}}}
\nc{\FF}{{{\mathbb F}}}
\nc{\NN}{{{\mathbb N}}}
\nc{\ZZ}{{{\mathbb Z}}}
\nc{\QQ}{{{\mathbb Q}}}
\nc{\UU}{{{\mathbb U}}}
\nc{\EE}{{{\mathbb E}}}

\nc{\bH}{{\mathfrak{H}}}

\nc{\sK}{{{\mathscr{K}}}}
\nc{\sS}{{{\mathscr{S}}}}
\nc{\sT}{{{\mathscr{T}}}}
\nc{\sA}{{{\mathscr{A}}}}
\nc{\sB}{{{\mathscr{B}}}}
\nc{\sC}{{{\mathscr{C}}}}
\nc{\sE}{{{\mathscr{E}}}}
\nc{\sL}{{{\mathscr{L}}}}
\nc{\sG}{{{\mathscr{G}}}}
\nc{\sF}{{{\mathscr{F}}}}
\nc{\sI}{{{\mathscr{I}}}}
\nc{\sN}{{{\mathscr{N}}}}
\nc{\sM}{{{\mathscr{M}}}}

\nc{\Choi}{Choi-Jamio\l{}kowski }
\nc{\reg}{\infty}
\nc{\mx}{\text{\rm mx}}
\nc{\amo}{\text{\rm amo}}
\nc{\Renyi}{R\'{e}nyi }

\nc{\conv}{\operatorname{conv}}
\nc{\cvxset}{\mathscr{C}}

\nc{\RM}{{{\mathscr{R}}}}

\nc{\END}{\operatorname{End}}
\nc{\PERM}{\mathfrak{\sigma}}

\nc{\Cone}{\text{\rm Cone}}
\nc{\sep}{{\SEP}}

\nc{\DD}{{{\mathbb D}}}
\nc{\BS}{{\scriptscriptstyle \rm {BS}}}
\nc{\Sand}{{\scriptscriptstyle  \rm S}}
\nc{\Petz}{{\scriptscriptstyle  \rm P}}
\nc{\Hypo}{{\scriptscriptstyle  \rm H}}
\nc{\Meas}{{\scriptscriptstyle \rm M}}
\nc{\Proj}{{{\scriptscriptstyle \rm P}}}
\nc{\KL}{{{\scriptscriptstyle \rm KL}}}
\nc{\IS}{{{\scriptscriptstyle \rm IS}}}

\nc{\suchthat}{\text{\rm s.t.}}

\nc{\pl}{{\scalebox{0.7}{+}}}
\nc{\HERM}{\mathscr{H}}
\nc{\PSD}{\HERM_{\pl}}
\nc{\PD}{\HERM_{\pl\pl}}
\nc{\density}{\mathscr{D}}
\nc{\subdensity}{\mathscr{D}_\bullet}

\nc{\polarPSD}[1]{{#1}_{\pl}^{\circ}}
\nc{\polarPSDre}[1]{{#1}_{\pl}^{\star}}
\nc{\polarPD}[1]{{#1}_{\pl\pl}^{\circ}}

\nc{\PPT}{\text{\rm PPT}}
\nc{\Rains}{\text{\rm Rains}}
\nc{\WD}{\text{\rm WD}}
\nc{\SEP}{\text{\rm SEP}}
\nc{\PSEP}{\text{\rm PSEP}}
\nc{\CPTP}{\text{\rm CPTP}}
\nc{\POVM}{\text{\rm POVM}}
\nc{\PVM}{\text{\rm PVM}}
\nc{\CP}{\text{\rm CP}}
\nc{\adv}{\text{\rm adv}}
\nc{\spec}{\text{\rm spec}}
\nc{\poly}{\text{\rm poly}}
\nc{\End}{\operatorname{End}}
\nc{\Par}{\operatorname{Par}}
\nc{\RNG}{\operatorname{RNG}}
\newcommand{\STAB}{\operatorname{Stab}}
\newcommand{\All}{\operatorname{All}}
\nc{\epi}{\boldsymbol{\operatorname{epi}}}
\nc{\op}{\boldsymbol{\operatorname{op}}}

\makeatletter
\newcommand*\rel@kern[1]{\kern#1\dimexpr\macc@kerna}
\newcommand*\widebar[1]{%
  \begingroup
  \def\mathaccent##1##2{%
    \rel@kern{0.8}%
    \overline{\rel@kern{-0.8}\macc@nucleus\rel@kern{0.2}}%
    \rel@kern{-0.2}%
  }%
  \macc@depth\@ne
  \let\math@bgroup\@empty \let\math@egroup\macc@set@skewchar
  \mathsurround\z@ \frozen@everymath{\mathgroup\macc@group\relax}%
  \macc@set@skewchar\relax
  \let\mathaccentV\macc@nested@a
  \macc@nested@a\relax111{#1}%
  \endgroup
}
\makeatother

\usepackage[T1]{fontenc}
\usepackage[utf8]{inputenc}
\usepackage[normalem]{ulem}

\newcommand{\Fd}{\mathbb F_d}
\newcommand{\Stab}{\mathsf S}
\newcommand{\Amb}{a_{\Stab}}
\newcommand{\cEnc}{\mathcal C}
\newcommand{\cEra}{\mathcal E}

\newcommand{\cRec}{\mathcal R}
\newcommand{\Dbin}{D_{\mathrm{bin}}}

\begin{document}

\title{\Large \textbf{Reliability of quantum communication over erasure channels}}
\author[1]{Hongyu Mao}
\author[1]{Kun Fang~\thanks{kunfang@cuhk.edu.cn}}
\affil[1]{\small School of Data Science, The Chinese University of Hong Kong, Shenzhen\protect \\
Guangdong, 518172, China}
\date{November 18, 2026}
\maketitle

\begin{abstract}
Quantum information can be reliably transmitted at rates below the quantum capacity with an error that vanishes as the number of channel uses grows. The reliability function characterizes the optimal exponential decay rate of this error. For quantum erasure channels with a prime input dimension, we determine the reliability function exactly at all positive communication rates below capacity that meet or exceed an explicit critical rate. Notably, stabilizer coding strategies attain this optimal reliability throughout this regime.
\end{abstract}

\tableofcontents

\section{Introduction}\label{sec:background}

\subsection{Background}

Quantum communication transmits quantum states and distributes
entanglement between distant systems, enabling remote quantum
processors to perform networked quantum information processing
\cite{kimbleQuantumInternet2008}.
A communication protocol encodes logical quantum systems into
physical carriers, sends them through a quantum channel, and applies
a recovery operation at the receiver.  We consider unassisted
entanglement transmission: the sender transmits one half of a
maximally entangled state while retaining the other half, without
preshared entanglement, feedback, or auxiliary classical communication \cite{schumacherSendingEntanglementNoisy1996,barnumInformationTransmissionNoisy1998a}.
We study this task over the quantum erasure channel, which replaces
a transmitted system by an orthogonal flag with probability $p$,
thereby revealing the erased positions to the receiver
\cite{bennettCapacitiesQuantumErasure1997}.
Such detectable information loss can arise, for example, from atom
loss or leakage out of the computational subspace in neutral-atom processors
\cite{wuErasureConversion2022,chowCircuitBasedLeakageErasure2024}.
The independent erasure model isolates this mechanism by assuming
erasures that are independent of the input state and across channel
uses, perfectly known erasure locations, and noiseless surviving systems.

The quantum capacity is the supremum of rates at which entanglement
can be transmitted with error tending to zero as the blocklength,
the number of channel uses, grows.  Here the entanglement fidelity
$F$ is the overlap of the recovered joint state with the maximally
entangled target, and the error is the entanglement infidelity $1-F$.
For systems of local Hilbert-space dimension $d$, the erasure
channel has quantum capacity $\max\{1-2p,0\}\ln d$ nats per channel
use \cite{bennettCapacitiesQuantumErasure1997}; the information rate
is the natural logarithm of the transmitted logical dimension divided
by the blocklength.  Capacity specifies the largest rate permitting
vanishing error, but does not specify how rapidly the error vanishes.
At a fixed rate below capacity, the reliability function quantifies
the exponential decay of the optimal entanglement infidelity
\cite[Section~II]{bargLowrateBoundReliability2002}.
It thus measures the asymptotic blocklength cost of suppressing error
and distinguishes schemes that approach unit fidelity at different
speeds while operating at the same rate.

A stabilizer coding scheme combines a stabilizer encoder with syndrome
correction.  The encoder maps logical quantum systems into a common
eigenspace of commuting generalized Pauli operators.  It can be implemented
by a Clifford circuit, whose gates map Pauli operators to Pauli operators
under conjugation \cite[Chapters~3--4]{gottesmanStabilizerCodes1997}.
At the receiver, syndrome correction measures the stabilizer operators;
their outcomes form a syndrome that guides a conditional Pauli correction
without measuring the encoded logical information.  The receiver then
undoes the Clifford encoding to recover the logical systems
\cite[Chapter~3]{gottesmanStabilizerCodes1997}. When the local
dimension $d$ is prime, Pauli operators can be labeled by vectors over the
finite field $\mathbb F_d$, expressing commutation and syndrome constraints
as linear algebra over that field
\cite[Section~2]{gottesmanHigherDimensional1998}.  The resulting scheme
requires no non-Clifford operations for encoding and syndrome correction,
while its algebraic structure facilitates code construction and analysis.
These features motivate
achievability bounds based on stabilizer encoders and syndrome correction,
and raise the question of whether such schemes attain the optimal reliability
among arbitrary quantum encoders and recoveries.

Achievability bounds on quantum reliability have been developed through
random coding, which averages over an ensemble of codes, and
weight-spectrum methods, which count errors by the number of physical
systems on which they act.  Hamada used random stabilizer codes to
derive achievable error exponents for memoryless quantum channels,
whose successive uses act independently
\cite{hamadaLowerBoundsQuantum2002}.
Barg strengthened Hamada's random-coding lower bound through
expurgation for stabilizer codes.  He related stabilizer decoding
to a classical additive channel and combined constraints on the
error-label spectrum with bounds on the probability of confusing
two errors to derive an expurgated exponent.  This exponent exceeds
the random-coding exponent over a nonempty interval of low rates
when the noise is sufficiently weak, as illustrated for the
depolarizing channel
\cite[Section~V, Theorem~6]{bargLowrateBoundReliability2002}.
Ashikhmin bounded fidelity through weight spectra of stabilizer and
Calderbank--Shor--Steane (CSS) codes, a subclass of stabilizer codes
with separate Pauli shift and phase checks, and through expurgated
ensembles obtained by discarding codes with undesirable low-weight
errors \cite{ashikhminFidelityLowerBounds2014}.
These constructions provide achievable bounds within stabilizer
coding and hence lower bounds on the optimal reliability when
arbitrary encoders are permitted; they do not supply a converse for
that larger class.

Finite-blocklength bounds for the erasure channel provide related
results, but their implications for the optimal asymptotic error exponent
require further analysis.  For the qubit erasure channel, Ashikhmin gave
linear-programming lower bounds on unrecoverable erasure-set counts at
each size and on decoding error probability for stabilizer codes, as
well as achievability bounds for stabilizer and CSS codes
\cite{ashikhminFidelityFiniteLength2020}.  Tightness at a particular
blocklength alone does not determine the optimal exponent at a fixed
positive rate as the blocklength grows.
For a fixed erasure pattern, Gullans et al. characterized the optimal
recovery probability of a qubit stabilizer code through the number of
independent errors with zero syndrome and nontrivial logical action
\cite[Section~III.B, Eqs.~(17)--(18)]{gullansQuantumCodingLowdepth2021journal}.
This characterization makes explicit how ambiguity between logical actions
limits recovery.
Ostrev derived finite-blocklength achievability and converse bounds for
guessing the full physical Pauli error from its syndrome and side
information, including the erased positions
\cite[Definitions~2--3 and Theorems~7--8]{ostrevCanonicalFormFinite2026a}.
For erasures, his achievability bound yields the qubit random-coding
exponent considered here.  At sufficiently high rates below capacity,
this exponent matches that of his guessing converse, as follows from binomial
large-deviation estimates \cite[Section~5.4]{ostrevCanonicalFormFinite2026a}.
However, this converse applies to physical-error guessing: logical recovery
need only distinguish errors by their action on the encoded information.
It therefore does not by itself provide a converse for logical recovery or
arbitrary quantum encoding.  Other finite-blocklength studies consider
quantum polar codes \cite{duttonPerformancePolarCodes2012}.

These random-coding, expurgated, and finite-blocklength results
motivate a reliability upper bound for unassisted entanglement
transmission over the quantum erasure channel that applies to
arbitrary quantum encoders and recoveries and matches stabilizer
achievability.  This requires a converse for the entanglement
infidelity itself and identification of the rates where the bounds
coincide.  We establish such a converse and determine the exact
reliability function for positive rates at or above a critical rate
and below capacity, showing that stabilizer encoding with syndrome
correction attains
the unrestricted optimal error exponent throughout this region.

\subsection{Contributions}

Consider unassisted entanglement transmission over an independent
erasure channel with prime local dimension $d$, erasure probability
$0<p<1/2$, and erased positions known to the receiver.  An encoder maps
$k$ logical $d$-dimensional systems to $n$ physical $d$-dimensional
systems, giving an actual rate $k/n$ per channel use; $r$ denotes the
target rate, so the codes considered have $k/n\ge r$.  In these normalized
units, the quantum capacity is the largest rate permitting vanishing error.
Entanglement fidelity is the recovered state's overlap
with a maximally entangled target.  Error is one minus this fidelity;
the reliability function gives its optimal exponential decay rate
with blocklength at a given communication rate.  Stabilizer codes
use a common eigenspace of commuting generalized Pauli operators,
whose measured eigenvalues form a correction syndrome.  Our contributions
are summarized in the following two points.

\begin{enumerate}
  \item For every positive rate below quantum capacity, we prove a
  reliability upper bound valid for arbitrary completely positive,
  trace-preserving (CPTP) encoders and
  recoveries.  The bound is explicitly computable from the rate and
  erasure probability: it is the binary relative entropy between
  the erasure fraction at the capacity boundary for that rate and
  the channel's actual erasure probability.  Binary relative entropy
  measures the exponential probability cost of observing the former
  erasure fraction under the latter probability.  A change of erasure
  probability transfers an above-capacity constant-error bound
  \cite[Theorem~2]{morganPrettyStrongConverse2014} to a converse for
  rates below capacity (Proposition~\ref{prop:reliability-converse}).
  We combine it with stabilizer encoding and syndrome decoding, using
  random coding and expurgation of
  codes containing nontrivial logical errors supported on few physical
  systems, developing the random-coding
  and spectrum methods of
  \cite{hamadaLowerBoundsQuantum2002,bargLowrateBoundReliability2002,ashikhminFidelityLowerBounds2014}
  into the bounds of Theorem~\ref{thm:reliability-main}
  (Propositions~\ref{thm:random-coding-reliability} and
  \ref{thm:logical-expurgated-reliability}).

  \item For positive rates below capacity and at or above the critical
  rate $[1-(d^2+1)p]/[1+(d^2-1)p]$, the random-coding bound
  (Proposition~\ref{thm:random-coding-reliability}) matches the converse
  for arbitrary encoders and recoveries.  The stabilizer and
  unrestricted reliability functions therefore exist and coincide: a
  stabilizer encoder and syndrome decoder attain the converse error exponent
  (Corollary~\ref{prop:reliability-matching}).  When $p\ge1/(d^2+1)$,
  this result holds at every positive rate below capacity.
\end{enumerate}

\subsection{Organization of the paper}\label{subsec:paper-organization}
Section~\ref{sec:reliability-definition} defines the communication task
and reliability functions, develops stabilizer structure and syndrome
correction, and establishes optimal recovery and the probabilistic
formulation used for achievability.  Section~\ref{sec:main-results} presents the random-coding
and expurgated lower bounds, the converse for arbitrary encoders and
recoveries, the rate region where the bounds match, and numerical
illustrations.  Section~\ref{sec:conclusion} concludes the paper.
Appendix~\ref{app:binomial-estimates} collects the binomial estimates
used in the proofs, while Appendix~\ref{app:ostrev} compares
entanglement transmission with physical-error guessing and
no-feedback classical communication.
\section{Reliability functions and stabilizer syndrome correction}

\label{sec:reliability-definition}

We first define the reliability functions for stabilizer and unrestricted
entanglement transmission.  We then develop the stabilizer structure
and its correction rules, determine optimal recovery for a fixed encoder,
and give a probabilistic expression for the stabilizer reliability functions.

Throughout the paper, $d$ is prime, $\mathbb F_d$ denotes the field with
$d$ elements, and $[n]:=\{1,\ldots,n\}$.
For finite-dimensional Hilbert spaces $\mathcal H$ and $\mathcal K$, let
$\mathcal L(\mathcal H)$ be the space of linear operators on $\mathcal H$
and $\mathrm{CPTP}(\mathcal H,\mathcal K)$ the set of CPTP maps from
$\mathcal L(\mathcal H)$ to $\mathcal L(\mathcal K)$.
All logarithms are natural.  Information rates are measured in nats
per channel use.

\subsection{Reliability functions for entanglement transmission}

\paragraph{Quantum communication protocol.}
We consider unassisted entanglement transmission over a quantum channel,
with no preshared entanglement, feedback, or auxiliary classical
communication.  Let $\mathcal H_A$ be an $M$-dimensional logical input
space and let $\mathcal H_R\cong\mathcal H_A$ be a reference space.
In orthonormal bases $\{|j\rangle_R\}_{j=1}^M$ and
$\{|j\rangle_A\}_{j=1}^M$, the maximally entangled state and its density
operator are
\begin{equation}
 |\Phi_M\rangle_{RA}:=\frac1{\sqrt M}\sum_{j=1}^M|j\rangle_R|j\rangle_A,
 \qquad \Phi_M:=|\Phi_M\rangle\langle\Phi_M|.
\end{equation}
The sender prepares $\Phi_M$ locally and retains $R$.  The input $A$ is
encoded into the channel input space $\mathcal H_P$ by a CPTP map
$\cEnc\in\mathrm{CPTP}(\mathcal H_A,\mathcal H_P)$, sent through a CPTP
channel $\mathcal N\in\mathrm{CPTP}(\mathcal H_P,\mathcal H_B)$, and
recovered by $\cRec\in\mathrm{CPTP}(\mathcal H_B,\mathcal H_A)$.
The spaces $\mathcal H_P$ and $\mathcal H_B$ need not coincide.

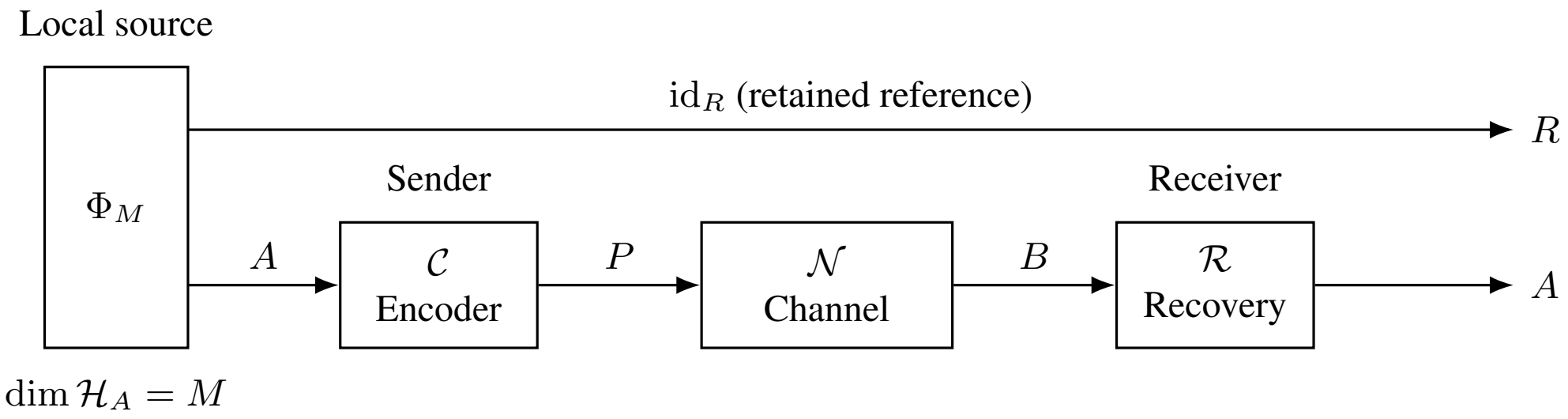
\begin{figure}[htbp]
 \centering
 \begin{tikzpicture}[
   font=\footnotesize,
   block/.style={draw, line width=0.6pt, minimum height=1.05cm,
                 minimum width=1.65cm, align=center},
   wire/.style={line width=0.6pt, -{Latex[length=2mm]}},
   x=1cm, y=1cm]
  \node[block, minimum width=1.2cm, minimum height=2.35cm]
    (source) at (0.8,0.65) {$\Phi_M$};
  \node[above] at (0.8,1.95) {Local source};
  \node[below] at (0.8,-0.65) {$\dim\mathcal H_A=M$};
  \node[block] (enc) at (3.5,0) {$\cEnc$\\Encoder};
  \node[block, minimum width=2.1cm] (era) at (6.75,0)
    {$\mathcal N$\\Channel};
  \node[block] (rec) at (10,0) {$\cRec$\\Recovery};
  \node[above] at (3.5,0.65) {Sender};
  \node[above] at (10,0.65) {Receiver};
  \draw[wire] (1.4,1.3) -- node[above] {$\id_R$ (retained reference)}
    (12.5,1.3) node[right] {$R$};
  \draw[wire] (1.4,0) -- node[above] {$A$} (enc.west);
  \draw[wire] (enc.east) -- node[above] {$P$} (era.west);
  \draw[wire] (era.east) -- node[above] {$B$} (rec.west);
  \draw[wire] (rec.east) -- (12.5,0) node[right] {$A$};
 \end{tikzpicture}
 \caption{Unassisted entanglement transmission over a channel
 $\mathcal N$.  The sender prepares $\Phi_M$ on $RA$, retains the
 reference $R$, and encodes $A$ into the channel input $P$.  The receiver
 applies the recovery $\cRec$ to the channel output $B$, producing a
 system with the same logical space $\mathcal H_A$ as the input.  The
 overlap of its joint state with $R$ and the target $\Phi_M$ is the
 entanglement fidelity.}
 \label{fig:entanglement-distribution}
\end{figure}

The entanglement fidelity is the overlap with the pure target, without
taking a square root:
\begin{equation}\label{eq:general-ent-fidelity}
 F(\cEnc,\cRec;\mathcal N)
 :=\langle\Phi_M|
 \bigl(\id_R\otimes(\cRec\circ\mathcal N\circ\cEnc)\bigr)(\Phi_M)
 |\Phi_M\rangle.
\end{equation}
Here $\id_R$ is the identity map on the reference system.  The
entanglement infidelity of the protocol is
$\varepsilon_{\rm ent}(\cEnc,\cRec;\mathcal N)
:=1-F(\cEnc,\cRec;\mathcal N)$.  Figure~\ref{fig:entanglement-distribution}
illustrates the protocol.

\paragraph{Quantum erasure channel.}
The quantum erasure channel replaces a qudit by an orthogonal erasure
flag $|e\rangle$ with probability $p$ and otherwise transmits it unchanged.Formally, the channel is the map
$\cEra_{p}:\mathcal L(\mathbb C^d)\to
\mathcal L(\mathbb C^d\oplus\mathbb C|e\rangle)$,
where $\mathbb C|e\rangle$ is the one-dimensional span of
$|e\rangle$, orthogonal to the input space $\mathbb C^d$.
The channel acts as
\begin{equation}
 \cEra_{p}(\rho):=(1-p)\rho+p\tr(\rho)|e\rangle\langle e|,
 \qquad 0\le p\le1,
\end{equation}
where $\rho$ is an input density operator, a positive operator with
trace one, and $\tr$ denotes the trace.

For $n\ge1$ independent channel uses, let $E_p\subseteq[n]$ be the
random set of erased positions.  Each position is erased independently
with probability $p$, so
\begin{equation}
 \Pr(E_p=E)=p^{|E|}(1-p)^{n-|E|},\qquad E\subseteq[n],
\end{equation}
where $|E|$ is the number of elements of $E$.  For each
$E\subseteq[n]$, define the fixed-pattern channel by
\begin{equation}\label{eq:fixed-pattern-erasure-channel}
 \cEra_E(\rho)
 \coloneqq\tr_E(\rho)\otimes|e_E\rangle\langle e_E|,
 \qquad |e_E\rangle=|e\rangle^{\otimes|E|},
\end{equation}
where the partial trace $\tr_E$ discards the qudits at positions in
$E$, and tensor factors retain their original positions.  Thus
\begin{equation}
 \cEra_{p}^{\otimes n}
 =\sum_{E\subseteq[n]}\Pr(E_p=E)\cEra_E.
\end{equation}
The erasure flags reveal $E$ to the decoder, which may choose a recovery
operation for that fixed pattern.

For $n$ independent uses, we take the logical input $A$ to consist of
$k\in\{0,\ldots,n\}$ qudits.  The corresponding spaces and dimension are
\begin{equation}
 \mathcal H_A=(\mathbb C^d)^{\otimes k},\qquad
 \mathcal H_P=(\mathbb C^d)^{\otimes n},\qquad
 \mathcal H_B=(\mathbb C^d\oplus\mathbb C|e\rangle)^{\otimes n},
 \qquad M=d^k.
\end{equation}
The normalized rate is $k/n$ logical qudits per channel use, and the
information rate is $\ln M/n=(k/n)\ln d$ nats per channel use.  Below,
$r$ denotes a target normalized rate.  Specializing
\eqref{eq:general-ent-fidelity} to $\mathcal N=\cEra_p^{\otimes n}$,
write
\begin{equation}\label{eq:ent-fidelity}
 F(\cEnc,\cRec,p)
 :=F(\cEnc,\cRec;\cEra_p^{\otimes n})
 =\langle\Phi_M|
 \bigl(\id_R\otimes(\cRec\circ\cEra_p^{\otimes n}\circ\cEnc)\bigr)
 (\Phi_M)|\Phi_M\rangle,
\end{equation}
and denote the corresponding entanglement infidelity by
\begin{equation}\label{eq:erasure-ep-definiton}
 \varepsilon_{\rm ent}(\cEnc,\cRec,p):=1-F(\cEnc,\cRec,p).
\end{equation}
The encoder and recovery form an $(n,k,\varepsilon)$
entanglement-transmission protocol for the erasure channel when
$\varepsilon_{\rm ent}(\cEnc,\cRec,p)=\varepsilon$.  The recovery may
condition its operations on the revealed erasure pattern and its own
measurement outcomes.

For a revealed erasure pattern $E\subseteq[n]$, the conditional
entanglement infidelity of the same encoder and recovery is
\begin{equation}\label{eq:conditional-entanglement-infidelity}
 \varepsilon_{\mathrm{ent}}(\cEnc,\cRec\mid E)
 :=1-\langle\Phi_M|
 \bigl(\id_R\otimes(\cRec\circ\cEra_E\circ\cEnc)\bigr)(\Phi_M)
 |\Phi_M\rangle\in[0,1].
\end{equation}
The corresponding conditional entanglement fidelity is one minus this
quantity.  Averaging over $E_p$ gives
\begin{equation}\label{eq:conditional-error-decomposition}
 \varepsilon_{\mathrm{ent}}(\cEnc,\cRec,p)
 =\sum_{E\subseteq[n]}\Pr(E_p=E)\,
   \varepsilon_{\mathrm{ent}}(\cEnc,\cRec\mid E).
\end{equation}

\paragraph{Reliability functions.}\label{par:reliability-functions}
\label{subsec:all-code-definition}
For fixed $n$ and $k$, we distinguish encoder and recovery classes.
The first subscript denotes the encoder and the second the recovery:
$\STAB$ stands for stabilizer and $\All$ for all CPTP operations.
We consider the pairs
\begin{equation}
 (\alpha,\beta)\in\{(\STAB,\STAB),(\STAB,\All),(\All,\STAB),(\All,\All)\}.
\end{equation}

For the encoder, $\mathfrak C_{\All}(n,k)$ consists of all CPTP maps from the
$k$ logical qudits to the $n$ physical qudits.  The class $\mathfrak C_{\STAB}(n,k)$
consists of isometric encoders of stabilizer codes, which encode into a
common eigenspace of commuting generalized Pauli operators; their
structure is described in Section~\ref{subsec:stabilizer-structure}.
For the recovery, $\mathfrak R_{\All}(n,k)$ is the class of all CPTP maps from
the flagged output space $\mathcal H_B$ to the logical space $\mathcal H_A$.
The class $\mathfrak R_{\STAB}(n,k)$ consists of recoveries implemented by stabilizer
operations, as specified in Section~\ref{subsec:stabilizer-structure};
it includes syndrome recovery.  We suppress $(n,k)$ when the dimensions are fixed.

For a target normalized rate $0\le r\le1$, define the optimal entanglement
infidelity by optimizing over the admissible numbers of logical qudits,
encoders, and recoveries:
\let\rateconstraintlabel\label
\begin{equation}\label{eq:all-code-optimal-error}\rateconstraintlabel{eq:ss-optimal-error}
 \varepsilon_{n,\alpha,\beta}^*(r,p;d)
 :=\min_{\substack{k\in\{0,\ldots,n\}\\ k/n\ge r}}
 \inf_{\substack{\cEnc\in\mathfrak C_{\alpha}(n,k)\\
                    \cRec\in\mathfrak R_{\beta}(n,k)}}
 \varepsilon_{\mathrm{ent}}(\cEnc,\cRec,p).
\end{equation}
Following the
reliability-function formulation of
\cite[Section~II]{bargLowrateBoundReliability2002}, define
\begin{align}
 \underline E_{\alpha,\beta}(r,p;d)
 &:=\liminf_{n\to\infty}-\frac1n\ln\varepsilon_{n,\alpha,\beta}^*(r,p;d),
 \label{eq:liminf reliability-exponents varepsilon}\\
 \overline E_{\alpha,\beta}(r,p;d)
 &:=\limsup_{n\to\infty}-\frac1n\ln\varepsilon_{n,\alpha,\beta}^*(r,p;d).
 \label{eq:limsup reliability-exponents varepsilon}
\end{align}
We use $-\ln0=+\infty$.  When the lower and upper limits coincide, their
common value is denoted by
\begin{equation}\label{eq:all-code-reliability-exponents}
 E_{\alpha,\beta}(r,p;d):=\underline E_{\alpha,\beta}(r,p;d)
 =\overline E_{\alpha,\beta}(r,p;d).
\end{equation}

We suppress exponent arguments $(r,p;d)$ when they are clear from
context.  The inclusions of encoder and recovery classes imply
\begin{equation}
 \begin{aligned}
 \varepsilon_{n,\All,\All}^*
 &\le\varepsilon_{n,\STAB,\All}^*\le\varepsilon_{n,\STAB,\STAB}^*,\\
 \varepsilon_{n,\All,\All}^*
 &\le\varepsilon_{n,\All,\STAB}^*\le\varepsilon_{n,\STAB,\STAB}^*.
 \end{aligned}
\end{equation}
Taking normalized negative logarithms gives the reverse ordering:
\begin{equation}\label{eq:code-class-exponent-ordering}
 \begin{aligned}
 \underline E_{\STAB,\STAB}
 &\le\underline E_{\STAB,\All}\le\underline E_{\All,\All},\qquad
 \overline E_{\STAB,\STAB}
 &\le\overline E_{\STAB,\All}\le\overline E_{\All,\All},\\
 \underline E_{\STAB,\STAB}
 &\le\underline E_{\All,\STAB}\le\underline E_{\All,\All},\qquad
 \overline E_{\STAB,\STAB}
 &\le\overline E_{\All,\STAB}\le\overline E_{\All,\All}.
 \end{aligned}
\end{equation}
For the erasure channel, the inequalities comparing the $(\STAB,\STAB)$
and $(\STAB,\All)$ reliability functions are in fact equalities, as shown
in Section~\ref{subsec:stabilizer-reliability-probability}.
At $r=0$, the admissible $k=0$ code transmits a one-dimensional logical
system perfectly, so each optimal infidelity is zero and its lower and
upper reliability functions are $+\infty$.  For $p>0$, this value differs
from a limit taken as $r\downarrow0$ through positive rates.

\FloatBarrier
\subsection{Stabilizer encoder and syndrome correction}
\label{subsec:stabilizer-structure}

\paragraph{Stabilizer encoder.}
We give the algebraic construction of the stabilizer encoders
used above and describe their syndrome correction rules.
Fix a nontrivial additive character $\chi:\Fd\to\mathbb C$:
$|\chi(a)|=1$, $\chi(a+b)=\chi(a)\chi(b)$, and $\chi$ is not
identically one.  For prime $d$, one may take
$\chi(a)=\exp(2\pi i a/d)$.  In the computational basis
$\{|j\rangle:j\in\Fd\}$, define the shift, phase, and Weyl operators
for $a,b,j\in\Fd$ and $u=(x,z)\in\Fd^{2n}$ by
\begin{align}
 X(a)|j\rangle&=|j+a\rangle,
 &Z(b)|j\rangle&=\chi(bj)|j\rangle,\\
 W_{a,b}&=X(a)Z(b),
 &W_u&=\bigotimes_{i=1}^n W_{x_i,z_i}.
\end{align}
The Weyl label $u=(x,z)$ records the shift and phase
parameters at each physical position, omitting global phases.
On $\Fd^{2n}$, define the symplectic form and the symplectic
orthogonal complement of any subspace $L\subseteq\Fd^{2n}$ by
\begin{align}
 \langle(x,z),(x',z')\rangle_{\rm sp}
 &:=x\cdot z'-z\cdot x',\\
 L^\perp
 &:=\{u\in\Fd^{2n}:\langle u,v\rangle_{\rm sp}=0
             \text{ for every }v\in L\}.
 \label{eq:symplectic-complement}
\end{align}
Here $x\cdot z'=\sum_{i=1}^n x_i z_i'$ is the dot product over
$\Fd$.  The symplectic form is an alternating, nondegenerate bilinear
pairing: each vector pairs to zero with itself, and only the zero
vector pairs to zero with every vector.  Its orthogonal complement
therefore describes the labels commuting with a given set of Weyl
operators.
Since $Z(b)X(a)=\chi(ab)X(a)Z(b)$, multiplication at each
coordinate gives the Weyl commutation relation
$W_uW_v=\chi(-\langle u,v\rangle_{\rm sp})W_vW_u$;
see also \cite[Section~2]{gottesmanHigherDimensional1998}.

To define an $[[n,k]]_d$ stabilizer code, fix a subspace
$\Stab\subseteq\Fd^{2n}$ of dimension $n-k$ that is isotropic,
meaning that $\langle s,t\rangle_{\rm sp}=0$ for all $s,t\in\Stab$.
By \eqref{eq:symplectic-complement}, each $s\in\Stab$ therefore lies
in $\Stab^\perp$, so $\Stab\subseteq\Stab^\perp$.
The Weyl commutation relation then implies that the operators
$W_s$, $s\in\Stab$, commute.

Choose a phase multiple $T_s$ of each $W_s$ so that the resulting
stabilizer operators form a group satisfying
\begin{equation}
 T_sT_{s'}=T_{s+s'}.
\end{equation}
These operators can be constructed by normalizing the Weyl operators
associated with a basis of $\Stab$ to have order $d$ and taking their
products.  Independence of the basis labels ensures that $T_s$ is scalar
only when $s=0$.  The associated stabilizer code is their common
$+1$ eigenspace,
\begin{equation}
 \mathcal H_{\Stab}
 :=\{|\psi\rangle\in\mathcal H_P:
       T_s|\psi\rangle=|\psi\rangle\text{ for all }s\in\Stab\},
 \qquad \dim\mathcal H_{\Stab}=d^k.
\end{equation}
Indeed, $d^{-(n-k)}\sum_{s\in\Stab}T_s$ is the projector onto
$\mathcal H_{\Stab}$, and only the identity term has nonzero trace,
giving the stated dimension.
By the commutation relation, $u\in\Stab^\perp$ exactly when
$W_u$ commutes with every stabilizer operator $T_s$.
A Clifford unitary is a unitary that sends each Weyl operator to
another Weyl operator up to phase under conjugation.  A Clifford
encoder applies such a unitary to the logical input and $n-k$ ancillas,
additional qudits initialized in a fixed pure state.
The prime-dimensional Clifford construction supplies a Clifford unitary
$U_{\Stab}$ and an encoding isometry $V$ into $\mathcal H_{\Stab}$
\cite[Sections~2--3]{gottesmanHigherDimensional1998}.  Taking the ancillas
to be initialized in $|0\rangle^{\otimes(n-k)}$, write
\begin{equation}\label{eq:clifford-stabilizer-encoder}
 V|\psi\rangle=U_{\Stab}\bigl(|\psi\rangle\otimes
 |0\rangle^{\otimes(n-k)}\bigr),\qquad
 \cEnc_{\Stab}(\rho)=V\rho V^\dagger.
\end{equation}
The encoder $\cEnc_{\Stab}$ used in the construction below is this
Clifford isometry.  The encoder class $\mathfrak C_{\STAB}$ also allows other
isometries onto the same stabilizer code space.

\paragraph{Syndrome correction.}\label{par:syndrome-correction}
For an encoding of $k$ logical qudits into $n$ physical qudits, the class
$\mathfrak R_{\STAB}(n,k)$ consists of CPTP recoveries from the flagged
output space $\mathcal H_B$ to the logical space $\mathcal H_A$ implemented
using the following operations:
\begin{itemize}
 \item preparation of stabilizer ancillas and discarding of systems;
 \item Clifford operations;
 \item generalized Pauli measurements;
 \item unrestricted classical processing and adaptive control at the receiver.
\end{itemize}
The receiver may use erasure flags and measurement outcomes locally,
with no classical communication between sender and receiver.
A recovery may replace erased systems before applying these operations.
Syndrome recovery preprocesses the output, measures the stabilizer,
applies a conditional Weyl correction, and undoes the Clifford encoding.

The preprocessing uses operations admitted in $\mathfrak R_{\STAB}(n,k)$, including
any permitted channel-specific interface operations.  It produces $n$
physical qudits and may retain a classical outcome $y$.  Denoting the
completely positive map associated with $y$ by $\mathcal P_y$, the output
and normalization are
\begin{equation}
 \mathcal P_y:\mathcal L(\mathcal H_B)\longrightarrow
 \mathcal L(\mathcal H_P),\qquad
 \sum_y\mathcal P_y\text{ is CPTP}.
\end{equation}
Thus $\tr\mathcal P_y(\tau)$ is the probability of $y$ for the channel
output $\tau$.  The complete recovery specifies an operation for every
possible outcome.  A preprocessing map exists only when the allowed
output interface supports its implementation; its existence alone does
not imply that the channel output's information is preserved.

\begin{figure}[htbp]
 \centering
 \begin{tikzpicture}[
   font=\footnotesize,
   block/.style={draw, line width=0.6pt, minimum height=0.95cm,
                 minimum width=1.48cm, align=center},
   wire/.style={line width=0.6pt, -{Latex[length=2mm]}},
   classical/.style={wire, dashed},
   x=1cm, y=1cm]
  \node[block] (enc) at (0.9,0) {$\cEnc_{\Stab}$\\Encoder};
  \node[block] (noise) at (3.15,0) {$\mathcal N$\\Channel};
  \node[block] (pre) at (5.4,0) {Preprocess\\$\mathcal P_y$};
  \node[block] (syn) at (7.65,0) {Measure\\syndrome};
  \node[block] (corr) at (9.9,0) {$W_{c(y,t)}^\dagger$\\Correction};
  \node[block] (undo) at (12.15,0) {Inverse\\encoding};
  \draw[wire] (-0.3,0) node[left] {$\rho$} -- (enc.west);
  \draw[wire] (enc.east) -- node[above] {$P$} (noise.west);
  \draw[wire] (noise.east) -- node[above] {$B$} (pre.west);
  \draw[wire] (pre.east) -- node[above] {$P$} (syn.west);
  \draw[wire] (syn.east) -- (corr.west);
  \draw[wire] (corr.east) -- (undo.west);
  \draw[wire] (undo.east) -- (13.35,0) node[right] {$A$};
  \draw[classical] (pre.north) -- ++(0,1.5) -|
    node[pos=0.25,above] {$y$} (corr.north);
  \draw[classical] (syn.north) -- ++(0,0.8) -|
    node[pos=0.25,above] {$t$} (corr.north);
 \end{tikzpicture}
 \caption{A stabilizer encoder and syndrome recovery for a channel
 $\mathcal N$.  Preprocessing produces the physical qudits and retains
 classical information $y$; stabilizer measurement gives the syndrome $t$.
 The receiver applies the Weyl correction selected by $(y,t)$, undoes
 the Clifford encoder, and discards its ancillas.  Solid arrows carry
 quantum systems and dashed arrows carry classical information.}
 \label{fig:quantum-error-correction}
\end{figure}
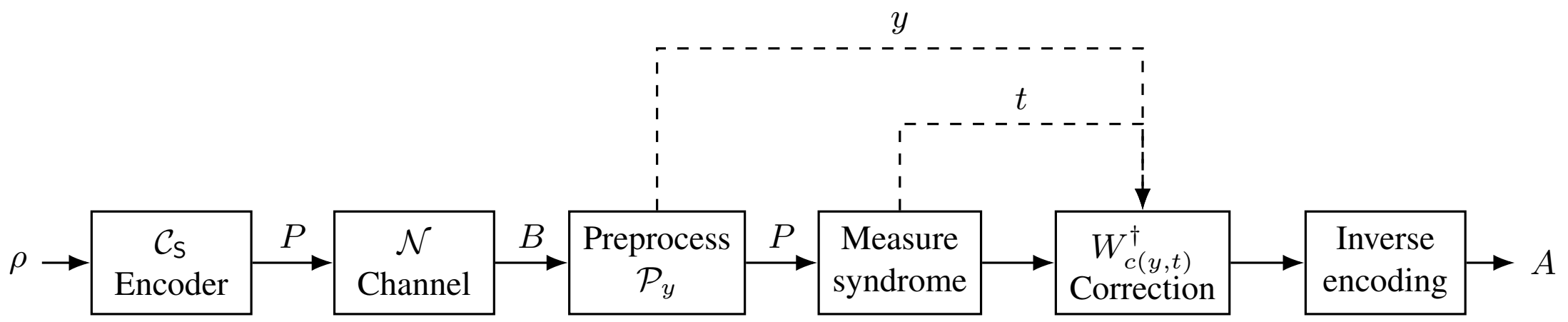

Fix an ordered basis $h_1,\ldots,h_{n-k}$ of $\Stab$, and measure the
corresponding commuting stabilizer generators.  For the outcome
$t=(t_1,\ldots,t_{n-k})\in\Fd^{n-k}$, let $\Pi_t$ be their joint
spectral projector, so that
\begin{equation}
 T_{h_i}\Pi_t=\chi(t_i)\Pi_t,\quad
 \Pi_t\Pi_{t'}=\delta_{t,t'}\Pi_t,\quad
 \sum_{t\in\Fd^{n-k}}\Pi_t=I_P.
\end{equation}
Here $\delta_{t,t'}$ is one if $t=t'$ and zero otherwise.  The measured
vector $t$ is the syndrome.  A classical correction rule assigns a Weyl
label $c(y,t)\in\Fd^{2n}$ to each syndrome $t$ and retained classical
outcome $y$.  After measuring $t$, the receiver applies
$W_{c(y,t)}^\dagger$.  For fixed $y$, the resulting channel is
\begin{equation}
 \mathcal Q_y(\tau):=\sum_{t\in\Fd^{n-k}}
 W_{c(y,t)}^\dagger\Pi_t\tau\Pi_t W_{c(y,t)},
 \qquad \tau\in\mathcal L(\mathcal H_P).
\end{equation}
The projectors sum to $I_P$ and the corrections are unitary, so
$\mathcal Q_y$ is CPTP.  The rule $c$ is classical and may depend on
side information; the measurement and Weyl correction are quantum
operations.  Applying $U_{\Stab}^\dagger$ and discarding the $n-k$
ancillas gives the inverse-encoding channel
\begin{equation}
 \mathcal V_{\rm inv}(\tau):=
 \tr_{\mathrm{anc}}\!\left(U_{\Stab}^\dagger\tau U_{\Stab}\right).
\end{equation}
Here $\tr_{\mathrm{anc}}$ denotes the partial trace over the $n-k$
ancilla systems of the encoder.
The resulting recovery from $\mathcal H_B$ to $\mathcal H_A$ is
\begin{equation}\label{eq:syndrome-recovery-map}
 \cRec=\sum_y\mathcal V_{\rm inv}\circ\mathcal Q_y\circ\mathcal P_y.
\end{equation}
The class $\mathfrak R_{\STAB}^{\mathrm{syn}}(n,k)$ contains these recoveries for
all permitted preprocessing instruments and classical correction rules.
Its dependence on the chosen Clifford encoder is suppressed in the notation.
Since these steps use only stabilizer operations,
$\mathfrak R_{\STAB}^{\mathrm{syn}}(n,k)\subseteq\mathfrak R_{\STAB}(n,k)$.
Figure~\ref{fig:quantum-error-correction} illustrates the three stages.

To see how the measurement guides the correction, consider the special case
where $\mathcal H_B=\mathcal H_P$, preprocessing is the identity, and the
channel applies a Weyl operator $W_u$, with $u\in\Fd^{2n}$.  To identify
the resulting syndrome, define the syndrome map
\begin{equation}\label{eq:stabilizer-syndrome-map}
 \sigma_{\Stab}:\Fd^{2n}\longrightarrow\Fd^{n-k},\qquad
 u\longmapsto\bigl(\langle u,h_i\rangle_{\rm sp}\bigr)_{i=1}^{n-k}.
\end{equation}
When $W_u$ acts on a code state, the measured syndrome is
$\sigma_{\Stab}(u)$ by the Weyl commutation relation.  Weyl operators
form a basis for operators on $\mathcal H_P$, so this relation can be
used to analyze general errors as well.  A linear combination of Weyl
operators can retain coherence between terms and does not by itself
specify a probability mixture of Weyl errors.

\subsection{Syndrome recovery for a fixed stabilizer encoder}
We now specialize the syndrome recovery to the quantum erasure channel.
The revealed erased set $E\subseteq[n]$ supplies the classical side
information.  For the syndrome recovery considered here, the receiver
replaces the erased systems by maximally mixed qudits, measures the stabilizer
syndrome, applies a Weyl correction selected from the erased set and
syndrome, and undoes the Clifford encoder.  We denote the class of these
recoveries, with the correction rule specified below, by
$\mathfrak R_{\STAB}^{\mathcal E}(n,k)$.  It is a subclass of the general
syndrome recoveries in \eqref{eq:syndrome-recovery-map}.   Together the classes satisfy
\begin{equation}\label{eq:recovery-class-inclusions}
 \mathfrak R_{\STAB}^{\mathcal E}(n,k)
 \subseteq\mathfrak R_{\STAB}^{\mathrm{syn}}(n,k)
 \subseteq\mathfrak R_{\STAB}(n,k)
 \subseteq\mathfrak R_{\All}(n,k).
\end{equation}
The first two classes depend on the chosen stabilizer code and its Clifford
encoder, although this dependence is suppressed in the notation.  The last
two classes are independent of the code and depend only on the dimensions.

\paragraph{Logical ambiguity.}
The Weyl labels supported on $E$ form
\begin{equation}
 V_E:=\{u=(x,z)\in\Fd^{2n}:x_i=z_i=0\text{ for every }i\notin E\}.
\end{equation}
Replacing the erased systems by locally prepared maximally mixed qudits
gives the channel
\begin{equation}
 \mathcal M_E(\rho):=\tr_E(\rho)\otimes\frac{I_E}{d^{|E|}},
\end{equation}
where $I_E$ is the identity on the qudits at positions in $E$, with
tensor factors placed at their original positions.  This replacement
uses only the known erasure positions and does not require access to the
lost systems.  Conversely, discarding the replacement qudits and
reinserting the erasure flags recovers $\cEra_E(\rho)$ as defined in
\eqref{eq:fixed-pattern-erasure-channel}.

The maximally mixed replacement is useful because $\mathcal M_E$ is a
uniform mixture of Weyl errors with labels in $V_E$.
Averaging over phase operators removes the off-diagonal entries
of a single-qudit operator $Y$, and averaging over shifts makes all
diagonal entries equal.  Hence
\begin{equation}
 \frac1{d^2}\sum_{a,b\in\Fd}W_{a,b}YW_{a,b}^\dagger
 =\tr(Y)\frac{I}{d}.
\end{equation}
Applying this identity at every position in $E$ gives
\begin{equation}\label{eq:erasure-weyl-twirl}
 \mathcal M_E(\rho)
 =\frac1{d^{2|E|}}\sum_{u\in V_E}W_u\rho W_u^\dagger.
\end{equation}

For each fixed erased set $E$, let $U_E$ be an error label drawn uniformly
from $V_E$:
\begin{equation}
 \Pr(U_E=u)
 =\begin{cases}
   d^{-2|E|},&u\in V_E,\\
   0,&u\notin V_E,
  \end{cases}
 \qquad u\in\Fd^{2n}.
\end{equation}
Thus
\begin{equation}
 \mathcal M_E(\rho)=\mathbb E\!\left[W_{U_E}\rho W_{U_E}^\dagger\right],
\end{equation}
where $\mathbb E$ denotes expectation over $U_E$.  This definition applies
to every $E$, including patterns with zero probability under $E_p$.  For
random erasures, the receiver knows $E_p$ when applying the recovery.

For the erasure channel, the retained preprocessing outcome is $y=E$,
so the correction rule is written $c(E,t)$.  Using the syndrome map
$\sigma_{\Stab}$ in \eqref{eq:stabilizer-syndrome-map}, we consider
recoveries whose correction label satisfies
\begin{equation}\label{eq:erasure-syndrome-correction}
 c(E,t)\in V_E,\qquad \sigma_{\Stab}(c(E,t))=t,
 \qquad t\in\sigma_{\Stab}(V_E).
\end{equation}
The label can be chosen separately for each erased set and attainable
syndrome; the choice need not be linear.  Such a label exists because
$t\in\sigma_{\Stab}(V_E)$.

More precisely, $\mathfrak R_{\STAB}^{\mathcal E}(n,k)$ consists of recoveries
that replace erased positions by maximally mixed qudits, measure the
syndrome, apply $W_{c(E,t)}^\dagger$ with a choice satisfying
\eqref{eq:erasure-syndrome-correction}, and undo the Clifford encoder
and discard its ancillas.  For unattainable syndromes, take $c(E,t)=0$.

We analyze the recovery for a fixed stabilizer encoder.  For an erased
set $E$, let $K_E$ be the zero-syndrome labels supported on $E$ and
$H_E$ the stabilizer labels supported on $E$.  Define their quotient and
its dimension by
\begin{equation}\label{eq:erasure-logical-ambiguity}
\begin{aligned}
 K_E&:=\Stab^\perp\cap V_E,
 &H_E&:=\Stab\cap V_E,\\
 G_E&:=K_E/H_E,
 &a_{\Stab}(E)&:=\dim_{\Fd}G_E.
\end{aligned}
\end{equation}
Here $H_E\subseteq K_E$ because $\Stab$ is isotropic, and $K_E/H_E$
denotes the quotient of their additive groups: its elements are the cosets
$u+H_E$ for $u\in K_E$.  It is also a vector space over $\Fd$.
For a given attainable syndrome, the difference between any two
compatible error labels lies in $K_E$.  Two such errors have the same
logical action exactly when their difference lies in $H_E$.  Therefore
$G_E$ describes their distinct relative logical actions, and
$|G_E|=d^{a_{\Stab}(E)}$ counts them.  We call $a_{\Stab}(E)$ the
logical ambiguity dimension.

\paragraph{Syndrome recovery.}
The counting argument agrees with the erasure-rank criterion in
\cite[Lemmas~1--2]{delfosseUpperBoundsRate2013journal} and with the qubit
logical-coset analysis of
\cite[Section~III.B, Eqs.~(17)--(18)]{gullansQuantumCodingLowdepth2021journal}.
Optimal Pauli recovery for a
qubit stabilizer encoding under general Pauli noise is established in
\cite[Appendix, Theorem~3]{fletcherStructuredNearoptimalChanneladapted2008}.
We give direct proofs for prime $d$, including a converse for arbitrary
CPTP recovery.

\begin{boxlemma}[Stabilizer recovery for a fixed stabilizer encoder]
\label{thm:fixed-encoder-stabilizer-recovery}\label{prop:syndrome-decoder-realization}
Let $\Stab$ be an $[[n,k]]_d$ stabilizer code with encoder $\cEnc_{\Stab}$ defined in
\eqref{eq:clifford-stabilizer-encoder}.  For each erased set $E$,
\begin{equation}\label{eq:fixed-pattern-stabilizer-recovery}
 \varepsilon_{\mathrm{ent}}(\cEnc_{\Stab},\cRec\mid E)
 =1-d^{-a_{\Stab}(E)},\qquad
 \forall\,\cRec\in\mathfrak R_{\STAB}^{\mathcal E}(n,k).
\end{equation}
\end{boxlemma}
\begin{proof}
Fix $\cRec\in\mathfrak R_{\STAB}^{\mathcal E}(n,k)$ and a choice of $c(E,t)$
implementing it.  We relate its conditional fidelity to logical
correction.  For each $E$, let $\cRec_E$ measure the stabilizer
syndrome $t$ on $\mathcal H_P$.  For an attainable outcome
$t\in\sigma_{\Stab}(V_E)$ it applies $W_{c(E,t)}^\dagger$,
undoes the Clifford encoder, and discards the encoding ancillas.
For any other outcome it applies the identity correction, undoes the
encoder, and discards the ancillas.  This defines a CPTP map from
$\mathcal H_P$ to $\mathcal H_A$.  The recovery $\cRec$
first measures the erasure flags, replaces the flagged positions by
maximally mixed qudits, and then applies
$\cRec_E$.  For an encoded state $V\rho V^\dagger$,
\eqref{eq:erasure-weyl-twirl} gives
\begin{align}
 (\cRec\circ\cEra_E)(V\rho V^\dagger)
 &=\cRec_E(\mathcal M_E(V\rho V^\dagger))\notag\\
 &=d^{-2|E|}\sum_{u\in V_E}
   \cRec_E
   (W_uV\rho V^\dagger W_u^\dagger).
   \label{eq:erasure-flag-to-weyl}
\end{align}

For the $u$ branch, the syndrome measurement returns
$t=\sigma_{\Stab}(u)$, so the recovery uses $c=c(E,t)$.
Since $\sigma_{\Stab}(c)=\sigma_{\Stab}(u)=t$, linearity gives
$\sigma_{\Stab}(u-c)=0$.  The kernel of $\sigma_{\Stab}$ is
$\Stab^\perp$ because the $h_i$ span $\Stab$.  The Weyl
multiplication rule also gives
\begin{equation}\label{eq:erasure-residual-logical-label}
 w:=u-c\in\ker\sigma_{\Stab}=\Stab^\perp,
 \qquad W_c^\dagger W_u=\omega_{u,c}W_w,
\end{equation}
where $|\omega_{u,c}|=1$.  The residual $w$ has zero syndrome but
need not lie in $\Stab$, so it can still cause a logical error.

On $\mathcal H_A$, define the logical Weyl operators by
$\overline W_{(a,b)}:=\bigotimes_{i=1}^kX(a_i)Z(b_i)$ for
$(a,b)\in\Fd^{2k}$.  The Clifford encoder identifies
$\Stab^\perp/\Stab$ with the logical Weyl labels
\cite[Section~2]{gottesmanHigherDimensional1998}: there is a linear map
\begin{equation}
 \lambda:\Stab^\perp\longrightarrow\Fd^{2k},\qquad
 w\longmapsto\lambda(w),\qquad \ker\lambda=\Stab,
 \qquad W_wV=e^{i\theta(w)}V\overline W_{\lambda(w)}.
\end{equation}
Thus applying the physical operator $W_w$ after encoding has the same
logical action as applying $\overline W_{\lambda(w)}$ before encoding,
up to a global phase.  When $\lambda(w)=0$, that logical action is
trivial.
Consequently the corrected term $(E,u)$ acts on the logical input,
up to a phase, as
$\overline W_{\ell(E,u)}$, where
\begin{equation}
 \ell(E,\cdot):V_E\longrightarrow\Fd^{2k},\qquad
 u\longmapsto\ell(E,u):=\lambda\!\left(
   u-c(E,\sigma_{\Stab}(u))\right).
\end{equation}
This residual-label map need not be linear, because the correction-label
choice need not be linear.

Since $\ker\lambda=\Stab$, the corrected logical action is trivial
exactly when $u-c(E,\sigma_{\Stab}(u))\in\Stab$.  As this residual
lies in $V_E$, this is equivalent to its membership in $H_E$.  In
particular, exact identification of the physical error is unnecessary.
Moreover, $H_E=K_E\cap\ker\lambda$, so $G_E$ identifies with a subspace
of the logical Weyl-label space $\Fd^{2k}$.

We count the successful labels.  On a fixed pattern
$E$, the restricted syndrome map
$\sigma_{\Stab}|_{V_E}:V_E\to\sigma_{\Stab}(V_E)$ has kernel $K_E$.
For any attainable syndrome $t$, exactly $|K_E|$ labels in $V_E$
have syndrome $t$, and $U_E$ is uniform over them conditional on that
syndrome.  For such a label
$u$, the residual $u-c(E,t)$ lies in $V_E$.  Correction succeeds
exactly when this residual lies in $\Stab$, hence in
$H_E=\Stab\cap V_E$.  The successful labels are therefore precisely
$c(E,t)+h$ for $h\in H_E$.  They all have syndrome $t$ because
$\sigma_{\Stab}(h)=0$, and there are $|H_E|$ of them.  The conditional
success probability is thus
\begin{equation}
 \Pr\!\left\{
   U_E-c(E,t)\in H_E
   \,\middle|\, \sigma_{\Stab}(U_E)=t
 \right\}
 =\frac{|H_E|}{|K_E|}=\frac1{|G_E|}=d^{-a_{\Stab}(E)}.
\end{equation}
Thus every such rule attains this value, independently of $t$.

Define
$|\Phi_\ell\rangle:=(I_R\otimes\overline W_\ell)|\Phi_M\rangle$.
With the Hilbert--Schmidt inner product
$\langle A,B\rangle_{\mathrm{HS}}:=\tr(A^\dagger B)$ on operators,
distinct logical Weyl operators are orthogonal:
a nonzero $X$ label has no diagonal entries, and a nonzero $Z$ label
has a vanishing character sum.  Therefore
\begin{equation}\label{eq:erasure-logical-bell-orthogonality}
 \langle\Phi_\ell|\Phi_{\ell'}\rangle
 =\frac1M\tr(\overline W_\ell^\dagger\overline W_{\ell'})
 =\delta_{\ell,\ell'},
 \qquad |\Phi_0\rangle=|\Phi_M\rangle.
\end{equation}
Here the Kronecker delta $\delta_{\ell,\ell'}$ equals one when
$\ell=\ell'$ and zero otherwise.
For each fixed $E$, the syndrome is deterministic in every Weyl term.
The correction leaves the reference untouched and sends $\Phi_M$ to
$|\Phi_{\ell(E,u)}\rangle\langle\Phi_{\ell(E,u)}|$ on the $u$ branch.
By \eqref{eq:erasure-logical-bell-orthogonality}, its overlap with the
target is one if $\ell(E,u)=0$ and zero otherwise.  Since
$\ker\lambda=\Stab$, the conditional entanglement fidelity
$F(\cEnc_{\Stab},\cRec\mid E)
:=1-\varepsilon_{\mathrm{ent}}(\cEnc_{\Stab},\cRec\mid E)$ is
\begin{equation}\label{eq:erasure-logical-decoding-error}
 F(\cEnc_{\Stab},\cRec\mid E)
 =\Pr\!\left\{
   U_E-c(E,\sigma_{\Stab}(U_E))\in\Stab\right\}.
\end{equation}
Combining the conditional fidelity identity with the count above gives
\eqref{eq:fixed-pattern-stabilizer-recovery}.
\end{proof}

We next bound the fidelity of an arbitrary CPTP recovery.

\begin{boxlemma}[Converse for a fixed stabilizer encoder]
\label{lem:fixed-encoder-recovery-converse}
Let $\Stab$ be an $[[n,k]]_d$ stabilizer code with the Clifford isometric
encoder $\cEnc_{\Stab}$ constructed in
Section~\ref{subsec:stabilizer-structure}.  For each erased set $E$,
\begin{equation}\label{eq:fixed-encoder-arbitrary-fidelity-bound}
 \varepsilon_{\mathrm{ent}}(\cEnc_{\Stab},\cRec\mid E)
 \ge 1-d^{-a_{\Stab}(E)},\qquad
 \forall\,\cRec\in\mathfrak R_{\All}(n,k).
\end{equation}
\end{boxlemma}
\begin{proof}
Fix an arbitrary CPTP recovery $\cRec$.  For each attainable
syndrome $t$, choose $v_t\in V_E$ with $\sigma_{\Stab}(v_t)=t$ and set
$J_t:=W_{v_t}V$.  The labels with syndrome $t$ form the set
$v_t+K_E$, which has probability
$p_t=|K_E|/|V_E|$.  For each $g\in G_E$, choose a representative
$k_g\in K_E$ and write $\overline W_g:=\overline W_{\lambda(k_g)}$.
Its Weyl label is independent of the representative, and the resulting logical
Weyl operators are distinct because
$K_E\cap\ker\lambda=H_E$.  The conditional logical
channel in the syndrome-$t$ sector is therefore
\begin{equation}
 \mathcal T_E(\rho)
 :=\frac1{|G_E|}\sum_{g\in G_E}
       \overline W_g\rho\overline W_g^\dagger,
\end{equation}
and the fixed-pattern channel has the block decomposition
\begin{equation}
 \mathcal M_E(V\rho V^\dagger)
 =\sum_{t\in\sigma_{\Stab}(V_E)}p_t
      J_t\mathcal T_E(\rho)J_t^\dagger.
\end{equation}
Indeed, each $u=v_t+k$ produces, up to a phase, the operator
$W_{v_t}W_kV=J_t\overline W_{\lambda(k)}$, while every coset of
$H_E$ in $K_E$ occurs $|H_E|$ times.  The ranges of the $J_t$ are
orthogonal syndrome sectors, and the $\overline W_g$ are distinct
logical Weyl operators, so
$\tr(\overline W_g^\dagger\overline W_h)=M\delta_{g,h}$.

Define the CPTP map on $\mathcal H_P$
$\cRec_E^{\mathrm{flag}}(\tau):=
\cRec(\tr_E(\tau)\otimes|e_E\rangle\langle e_E|)$, with tensor
factors returned to their original positions.  Since
$\cRec_E^{\mathrm{flag}}\circ\mathcal M_E
=\cRec\circ\cEra_E$, its restriction to sector $t$ induces
the logical CPTP map
$\mathcal Q_{E,t}(\rho):=
\cRec_E^{\mathrm{flag}}(J_t\rho J_t^\dagger)$.
For each $t$, let $Q_j$ be Kraus operators for $\mathcal Q_{E,t}$,
meaning that $\mathcal Q_{E,t}(\rho)=\sum_jQ_j\rho Q_j^\dagger$
and $\sum_jQ_j^\dagger Q_j=I$
\cite[Appendix]{fletcherStructuredNearoptimalChanneladapted2008}.
Suppress their dependence on $E$ and $t$, and
write $F_e(\mathcal N):=\langle\Phi_M|
(\id_R\otimes\mathcal N)(\Phi_M)|\Phi_M\rangle$ for a logical channel
$\mathcal N$.  Using the Hilbert--Schmidt inner product, we obtain
\begin{align}
 F_e(\cRec\circ\cEra_E\circ\cEnc_{\Stab})
 &=\sum_t p_tF_e(\mathcal Q_{E,t}\circ\mathcal T_E)\notag\\
 &=\frac1{|G_E|}\sum_t p_t\sum_{j,g}
   \left|\langle\Phi_M|
     (I_R\otimes Q_j\overline W_g)|\Phi_M\rangle\right|^2\notag\\
 &=\frac1{|G_E|M^2}\sum_t p_t\sum_{j,g}
      |\tr(Q_j\overline W_g)|^2\notag\\
 &=\frac1{|G_E|M}\sum_t p_t\sum_{j,g}
   \left|\left\langle
     \frac{\overline W_g^\dagger}{\sqrt M},Q_j
   \right\rangle_{\mathrm{HS}}\right|^2\notag\\
 &\le\frac1{|G_E|M}\sum_t p_t\sum_j\tr(Q_j^\dagger Q_j)\notag\\
 &=\frac{\tr(I_M)}{|G_E|M}\sum_t p_t\notag\\
 &=\frac1{|G_E|}=d^{-a_{\Stab}(E)}.
\end{align}
Here the first equality follows from the block decomposition,
$\cRec_E^{\mathrm{flag}}\circ\mathcal M_E=\cRec\circ\cEra_E$,
and linearity of entanglement fidelity.  The second uses the Kraus
operators $|G_E|^{-1/2}Q_j\overline W_g$ of
$\mathcal Q_{E,t}\circ\mathcal T_E$.  The third uses
$\langle\Phi_M|(I_R\otimes A)|\Phi_M\rangle=\tr(A)/M$.
The fourth follows from the Hilbert--Schmidt inner product and
cyclicity of the trace.  The inequality is Bessel's inequality for
the orthonormal operators $\overline W_g^\dagger/\sqrt M$, $g\in G_E$
\cite{melroseBessel2009}.
The Hilbert--Schmidt norm is
$\|A\|_{\mathrm{HS}}:=\sqrt{\tr(A^\dagger A)}$.
Explicitly, put
$A_g=\overline W_g^\dagger/\sqrt M$.  Orthogonality gives, for each $Q_j$,
\begin{equation}
 0\le\left\|Q_j-\sum_g\langle A_g,Q_j\rangle_{\mathrm{HS}}A_g
       \right\|_{\mathrm{HS}}^2
 =\tr(Q_j^\dagger Q_j)-\sum_g
      |\langle A_g,Q_j\rangle_{\mathrm{HS}}|^2,
\end{equation}
which is the inequality used in the fidelity calculation.
The remaining equalities use $\sum_jQ_j^\dagger Q_j=I_M$ for each $t$,
$\tr(I_M)=M$, $\sum_t p_t=1$, and $|G_E|=d^{a_{\Stab}(E)}$.
Taking one minus the fidelity proves the bound.
\end{proof}

The preceding two lemmas give matching bounds and hence determine
the optimal recovery for the Clifford encoder.

\begin{boxtheorem}[Optimal recovery for a fixed stabilizer encoder]
\label{prop:fixed-encoder-optimal-recovery}
\label{cor:syndrome-recovery-error-bound}
Let $\Stab$ be an $[[n,k]]_d$ stabilizer code with the Clifford isometric
encoder $\cEnc_{\Stab}$ constructed in
Section~\ref{subsec:stabilizer-structure}.  For each erased set $E$,
\begin{equation}\label{eq:fixed-pattern-optimal-fidelity}
 \inf_{\cRec\in\mathfrak R_{\STAB}(n,k)}
 \varepsilon_{\mathrm{ent}}(\cEnc_{\Stab},\cRec\mid E)
 =\inf_{\cRec\in\mathfrak R_{\All}(n,k)}
 \varepsilon_{\mathrm{ent}}(\cEnc_{\Stab},\cRec\mid E)
 =1-d^{-a_{\Stab}(E)}.
\end{equation}
For independent erasures with probability $p$ and erased set $E_p$,
\let\recoveryboundlabel\label
\begin{equation}\label{eq:fixed-encoder-optimal-error}\recoveryboundlabel{eq:fixed-encoder-stabilizer-recovery}
 \inf_{\cRec\in\mathfrak R_{\STAB}(n,k)}
 \varepsilon_{\mathrm{ent}}(\cEnc_{\Stab},\cRec,p)
 =\inf_{\cRec\in\mathfrak R_{\All}(n,k)}
 \varepsilon_{\mathrm{ent}}(\cEnc_{\Stab},\cRec,p)
 =\mathbb E_{E_p}\!\left[1-d^{-a_{\Stab}(E_p)}\right].
\end{equation}
\end{boxtheorem}
\begin{proof}
For a fixed $E$, the inclusions in \eqref{eq:recovery-class-inclusions},
Lemma~\ref{thm:fixed-encoder-stabilizer-recovery}, and
Lemma~\ref{lem:fixed-encoder-recovery-converse} give
\begin{align}
 1-d^{-a_{\Stab}(E)}
 &\le \inf_{\cRec\in\mathfrak R_{\All}(n,k)}
 \varepsilon_{\mathrm{ent}}(\cEnc_{\Stab},\cRec\mid E)\notag\\
 &\le \inf_{\cRec\in\mathfrak R_{\STAB}(n,k)}
 \varepsilon_{\mathrm{ent}}(\cEnc_{\Stab},\cRec\mid E)\notag\\
 &\le \inf_{\cRec\in\mathfrak R_{\STAB}^{\mathcal E}(n,k)}
 \varepsilon_{\mathrm{ent}}(\cEnc_{\Stab},\cRec\mid E)
 =1-d^{-a_{\Stab}(E)}.
\end{align}
The class $\mathfrak R_{\STAB}^{\mathcal E}(n,k)$ is nonempty, and every recovery
in it attains the displayed value.  The two ends of the chain agree, so all
terms are equal, proving \eqref{eq:fixed-pattern-optimal-fidelity}.

The erasure flags let each recovery in $\mathfrak R_{\STAB}^{\mathcal E}(n,k)$
select the appropriate correction for every pattern.  Averaging the
fixed-pattern infidelity therefore gives the right-hand side of
\eqref{eq:fixed-encoder-optimal-error} for each such recovery.  Conversely,
averaging the fixed-pattern converse gives the same lower bound for every
CPTP recovery.  Since $\mathfrak R_{\STAB}^{\mathcal E}(n,k)$ is a nonempty subset
of both recovery classes in \eqref{eq:fixed-encoder-optimal-error}, their
infima equal this value.
\end{proof}
\subsection{Equivalent probabilistic formulations of stabilizer reliability functions}
\label{subsec:stabilizer-reliability-probability}

\paragraph{Reduction to the minimum logical dimension.}
Write $\Stab:[[n,k]]_d$ when $\Stab$ is the stabilizer subspace of an
$[[n,k]]_d$ code.
By Theorem~\ref{prop:fixed-encoder-optimal-recovery} and definition \eqref{eq:all-code-optimal-error},
\begin{align}
 \varepsilon_{n,\STAB,\STAB}^*(r,p;d)
 &\coloneqq \min_{\substack{k\in\{0,\ldots,n\}\\ k/n\ge r}}
 \inf_{\substack{\cEnc\in\mathfrak C_{\STAB}(n,k)\\
                    \cRec\in\mathfrak R_{\STAB}(n,k)}}
 \varepsilon_{\mathrm{ent}}(\cEnc,\cRec,p) \notag\\
 &= \min_{\substack{k\in\{0,\ldots,n\}\\ k/n\ge r}}
 \inf_{\cEnc\in\mathfrak C_{\STAB}(n,k)}
 \inf_{\cRec\in\mathfrak R_{\STAB}(n,k)}
 \varepsilon_{\mathrm{ent}}(\cEnc,\cRec,p)\notag\\
 &= \min_{\substack{k\in\{0,\ldots,n\}\\ k/n\ge r}}
 \min_{\Stab:[[n,k]]_d}
 \inf_{\cRec\in\mathfrak R_{\STAB}(n,k)}
 \varepsilon_{\mathrm{ent}}(\cEnc_{\Stab},\cRec,p)\notag\\
 &= \min_{\substack{k\in\{0,\ldots,n\}\\ k/n\ge r}}
 \min_{\Stab:[[n,k]]_d}
 \mathbb E_{E_p}\!\left[1-d^{-a_{\Stab}(E_p)}\right].
 \label{eq:ss-optimization-expanded}
\end{align}
For the equality from the second to the third line, isometric encoders of
each code space differ by a logical unitary, which an arbitrary recovery can
undo; the Clifford encoder with syndrome recovery attains the resulting
optimal error.
The minimum exists because there are finitely many stabilizer subspaces.
The same $\Stab$ must be chosen before the erasure
pattern is known; the values $a_{\Stab}(E)$ cannot be minimized
independently for each $E$.  The same argument for arbitrary recovery in Theorem~\ref{prop:fixed-encoder-optimal-recovery} gives
\begin{equation}
 \varepsilon_{n,\STAB,\STAB}^*(r,p;d)
 =\varepsilon_{n,\STAB,\All}^*(r,p;d)
\end{equation}

The following lemma shows that the minimum is attained at the smallest
admissible logical dimension.
\begin{boxlemma}[Reduction to the minimum logical dimension]
\label{lem:minimum-logical-dimension}
For $0\le p\le1$, minimizing entanglement infidelity over
stabilizer codes with $k/n\ge r$ gives the same value as minimizing
over $[[n,\lceil nr\rceil]]_d$ stabilizer codes:
\begin{equation}
 \varepsilon_{n,\STAB,\STAB}^*(r,p;d)
 =\min_{\substack{\Stab:\ [[n,k]]_d\\ k=\lceil nr\rceil}}
   \mathbb E_{E_p}\!\left[1-d^{-a_{\Stab}(E_p)}\right].
 \label{eq:minimum-dimension-infidelity}
\end{equation}
Here $\lceil nr\rceil=nr$ whenever $nr$ is an integer.
For $p>0$, every code attaining the minimum entanglement infidelity
under the rate constraint has $k=\lceil nr\rceil$.
\end{boxlemma}
\begin{proof}
Let $\Stab$ be the stabilizer subspace of an $[[n,k]]_d$ code with
$k\ge \lceil nr\rceil$.  Every isotropic subspace of $\Fd^{2n}$ extends to an
isotropic subspace of dimension $n$: if its dimension is $m<n$,
its symplectic complement has dimension $2n-m>m$ and therefore
contains a vector outside the subspace.  Adjoining that vector
preserves isotropy because it is orthogonal to the subspace and
the symplectic form is alternating.  Repeating this process
as needed gives an isotropic extension
\begin{equation}
 \Stab\subseteq\Stab',\qquad \dim\Stab'=n-\lceil nr\rceil.
\end{equation}
Choosing the additional stabilizer eigenvalues consistently gives an
$[[n,\lceil nr\rceil]]_d$ subcode.  Since $(\Stab')^\perp\subseteq\Stab^\perp$,
for every $E\subseteq[n]$,
\begin{align}
 a_{\Stab'}(E)
 &=\dim((\Stab')^\perp\cap V_E)-\dim(\Stab'\cap V_E)\notag\\
 &\le\dim(\Stab^\perp\cap V_E)-\dim(\Stab\cap V_E)
 =a_{\Stab}(E).
\end{align}
Thus $1-d^{-a_{\Stab'}(E)}\le1-d^{-a_{\Stab}(E)}$ holds
pointwise.  Averaging shows that the expectation does not increase on
passing to the subcode.  For each code space, the optimized infidelity equals
this expectation by the preceding argument.  Because the $k=\lceil nr\rceil$
codes are also admissible under $k/n\ge r$, the two minimizations
have the same value.

If $k>\lceil nr\rceil$ and $p>0$, the all-erased pattern has probability $p^n>0$
and satisfies $a_{\Stab}([n])=2k$ and $a_{\Stab'}([n])=2\lceil nr\rceil$.
It follows that
\begin{align}
 \mathbb E_{E_p}\!\left[d^{-a_{\Stab'}(E_p)}
                         -d^{-a_{\Stab}(E_p)}\right]
 &\ge p^n\bigl(d^{-2\lceil nr\rceil}-d^{-2k}\bigr)>0.
\end{align}
For $p>0$, a code with $k>\lceil nr\rceil$ therefore cannot be optimal.
At $p=0$, every code space admits an encoder and recovery with zero
error, so the dimension of an optimal code need not equal $\lceil nr\rceil$.
\end{proof}

For $p>0$ and $r>0$, Theorem~\ref{prop:fixed-encoder-optimal-recovery}
and $a_{\Stab}([n])=2k$ give the all-erased lower bound
\begin{equation}
 \varepsilon_{n,\STAB,\STAB}^*(r,p;d)
 \ge p^n\bigl(1-d^{-2\lceil nr\rceil}\bigr)>0.
\end{equation}

Consequently, for $p,r>0$,
\begin{equation}
 \underline E_{\STAB,\STAB}(r,p;d)
 \le\overline E_{\STAB,\STAB}(r,p;d)\le-\ln p<+\infty,
\end{equation}
since $n^{-1}\ln(1-d^{-2\lceil nr\rceil})\to0$.
Thus, at positive erasure probability and positive rate, the optimal
infidelity cannot decay faster than exponentially in the blocklength:
the all-erased event alone imposes a finite upper bound on its decay
exponent.

\paragraph{Uncorrectable-pattern probability and equivalent reliability formulations.}
To analyze these reliability functions, it is convenient to use the
probability of an uncorrectable erasure pattern.
A pattern is exactly correctable if a recovery conditioned on that
pattern has entanglement fidelity one, equivalently zero infidelity.
For each fixed admissible stabilizer subspace $\Stab$,
\begin{equation}
 \text{$E$ is not exactly correctable}
 \ \Longleftrightarrow\ a_{\Stab}(E)>0
 \ \Longleftrightarrow\ G_E\ne\{0\}.
\end{equation}
Define the minimum uncorrectable-pattern probability by
\begin{equation}
 \eta_{n,\STAB,\STAB}^*(r,p;d)
 :=\min_{\substack{\Stab:\ [[n,k]]_d\\ k=\lceil nr\rceil}}
   \Pr_{E_p}\{a_{\Stab}(E_p)>0\}.\label{eq: def eta_n^*(r,p)}
\end{equation}
The extension in Lemma~\ref{lem:minimum-logical-dimension} cannot
increase $a_{\Stab}(E)$ for any $E$, so minimizing this probability over
all $k/n\ge r$ gives the same value.

The following lemma connects this probability to the stabilizer reliability functions.

\begin{boxlemma}[Equivalent formulations of the stabilizer reliability functions]
\label{lem:stabilizer-error-exponent-equivalence}
For fixed prime $d$, $0\le r\le1$, and $0\le p\le1$,
optimizing uncorrectable-pattern probability over stabilizer encoders
gives the same lower and upper reliability functions as optimizing entanglement
infidelity:
\begin{align}
 \underline E_{\STAB,\STAB}(r,p;d)
 &=\liminf_{n\to\infty}-\frac1n\ln\eta_{n,\STAB,\STAB}^*(r,p;d), \label{eq:liminf reliability-exponents eta}\\
 \overline E_{\STAB,\STAB}(r,p;d)
 &=\limsup_{n\to\infty}-\frac1n\ln\eta_{n,\STAB,\STAB}^*(r,p;d).\label{eq:limsup reliability-exponents eta}
\end{align}
\end{boxlemma}
\begin{proof}
Since $a_{\Stab}(E)$ is a nonnegative integer,
\begin{equation}
 (1-d^{-1})\mathbf 1\{a_{\Stab}(E)>0\}
 \le 1-d^{-a_{\Stab}(E)}
 \le \mathbf 1\{a_{\Stab}(E)>0\}.
\end{equation}
Taking expectations over the random erasure set $E_p$ gives, for every
stabilizer subspace $\Stab$,
\begin{equation}
 (1-d^{-1})\Pr_{E_p}\{a_{\Stab}(E_p)>0\}
 \le \mathbb E_{E_p}\!\left[1-d^{-a_{\Stab}(E_p)}\right]
 \le \Pr_{E_p}\{a_{\Stab}(E_p)>0\}.
\end{equation}
Minimizing all three terms over stabilizer subspaces with
$k=\lceil nr\rceil$ and applying
Lemma~\ref{lem:minimum-logical-dimension} to the middle term yields
\begin{equation}\label{eq: bound between eta and varepsilon}
 (1-d^{-1})\eta_{n,\STAB,\STAB}^*(r,p;d)
 \le \varepsilon_{n,\STAB,\STAB}^*(r,p;d)
 \le \eta_{n,\STAB,\STAB}^*(r,p;d).
\end{equation}

When $\eta_{n,\STAB,\STAB}^*(r,p;d)>0$,
\eqref{eq: bound between eta and varepsilon} also makes
$\varepsilon_{n,\STAB,\STAB}^*(r,p;d)>0$.  Taking negative logarithms
and dividing by $n$ gives, with the arguments $(r,p;d)$ suppressed,
\begin{equation}
 -\frac1n\ln\eta_{n,\STAB,\STAB}^*
 \le -\frac1n\ln\varepsilon_{n,\STAB,\STAB}^*
 \le -\frac1n\ln\eta_{n,\STAB,\STAB}^*-\frac1n\ln(1-d^{-1}).
\end{equation}
The two normalized negative logarithms differ by at most
$-n^{-1}\ln(1-d^{-1})$, which tends to zero because $d$ is fixed.
If $\eta_{n,\STAB,\STAB}^*(r,p;d)=0$, then the same finite-blocklength
bounds give $\varepsilon_{n,\STAB,\STAB}^*(r,p;d)=0$, and both
normalized negative logarithms equal $+\infty$.
Thus the inequalities hold in the extended real sense for every $n$.
Taking their lower and upper limits proves the two identities.
\end{proof}
\section{Reliability bounds and tightness}\label{sec:main-results}

The quantum capacity $Q$ is the supremum of information rates
achievable with entanglement infidelity tending to zero as the
blocklength grows.  For the erasure channel, in nats per channel use,
it is
\cite{bennettCapacitiesQuantumErasure1997}
\begin{equation}\label{eq:erasure-capacity}
 Q(\cEra_{p})=(1-2p)_+\ln d,
 \qquad x_+:=\max\{x,0\}.
\end{equation}
In the normalized rate units used here, dividing by $\ln d$ gives
capacity $1-2p$ for $0<p<1/2$.
Throughout this section, the bounds concern $0<p<1/2$ and normalized
rates $0<r<1-2p$.  For a candidate erasure
fraction $t$, the binary relative entropy measures its probability
cost relative to the channel erasure probability $p$.  Define
\begin{equation}\label{eq:binary-relative-entropy}
 \Dbin(t\|p):=t\ln\frac{t}{p}+(1-t)\ln\frac{1-t}{1-p},
 \qquad 0\le t\le1,
\end{equation}
with the continuous endpoint convention $0\ln0=0$.
Write $h(t)=-t\ln t-(1-t)\ln(1-t)$ for binary entropy, with
$h(0)=h(1)=0$.  The $d^2$-ary entropy is the entropy of a
distribution assigning mass $1-\delta$ to the zero Weyl label and
equal mass $\delta/(d^2-1)$ to each nonzero single-qudit label.
In nats it is
\begin{equation}\label{eq:q2-ary-entropy}
 H_{d^2}(\delta)
 :=-\delta\ln\delta-(1-\delta)\ln(1-\delta)
   +\delta\ln(d^2-1),\qquad 0\le\delta\le1.
\end{equation}
The logical weight spectrum counts labels with a nontrivial action
on the code by their weight, the number of physical positions where
their coordinate pair is nonzero.  Expurgation discards codes with
such labels below a selected weight threshold.
The Gilbert--Varshamov (GV) relative-distance threshold balances the
exponential number of low-weight labels against the number of
syndromes, as in the spectrum methods of
\cite{bargLowrateBoundReliability2002,ashikhminFidelityLowerBounds2014}.
For $0\le r<1$, define $\delta_{\rm GV}(r;d)$ as the unique solution in
$[0,1-d^{-2}]$ of
\begin{equation}\label{eq:logical-gv-distance}
 H_{d^2}(\delta_{\rm GV}(r;d))=(1-r)\ln d.
\end{equation}
This is a threshold used to select codes, rather than a distance
assigned to every stabilizer code.
The random-coding bound averages over uniformly selected stabilizer
label subspaces, with $t$ denoting the erased fraction.
The expurgated bound uses a normalized weight $\omega=w/n\in[0,1]$,
where $w$ is the number of nonzero coordinate pairs.
The corresponding lower bounds are
\begin{align}\label{eq:random-coding-exponent}
 E_{\rm rc}(r,p;d)
 &:=\min_{0\le t\le1}
 \left\{\Dbin(t\|p)+(1-r-2t)_+\ln d\right\}. \\
 E_{\rm expur}(r,p;d)
 &:={\left[(1-r)\ln d-
    \max_{\delta_{\rm GV}(r;d)\le\omega\le1}
       \{H_{d^2}(\omega)+\omega\ln p\}\right]}_+
 \label{eq:logical-expurgated-spectral}
\end{align}

We use the encoder and recovery classes and the encoder--recovery subscripts
defined in Section~\ref{subsec:all-code-definition}.

\begin{boxtheorem}[Bounds on the reliability functions]
\label{thm:reliability-main}
Let $d$ be prime, $0<p<1/2$, and $0<r<1-2p$.  Then
\begin{equation}\label{eq:reliability-main-bounds}
 \max\{E_{\rm rc},E_{\rm expur}\}
 \le\underline E_{\STAB,\STAB} \le\overline E_{\All,\All}
 \le\Dbin\!\left(\frac{1-r}{2}\middle\|p\right).
\end{equation}
The arguments $(r,p;d)$ of $E_{\rm rc}$, $E_{\rm expur}$ and reliability functions are suppressed.
\end{boxtheorem}

The random-coding and expurgation constructions in
Propositions~\ref{thm:random-coding-reliability} and
\ref{thm:logical-expurgated-reliability} bound the minimum probability of
an uncorrectable erasure pattern at $k=\lceil nr\rceil$.
Lemma~\ref{lem:stabilizer-error-exponent-equivalence} identifies its
lower and upper exponential rates with the corresponding stabilizer
reliability functions, giving the
first inequality in \eqref{eq:reliability-main-bounds}.  The underlying
optimal recovery is syndrome correction by
Theorem~\ref{prop:fixed-encoder-optimal-recovery}.
The inclusions of the recovery and encoder classes give
$\varepsilon_{n,\All,\All}^*\le\varepsilon_{n,\STAB,\STAB}^*$; taking normalized negative logarithms
and lower or upper limits yields the corresponding class orderings.
Finally, Proposition~\ref{prop:reliability-converse} bounds
$\overline E_{\All,\All}$ from above, completing the comparison.

When the random-coding lower bound meets the converse, these bounds
determine the reliability function exactly.  The following corollary
identifies this rate region and shows that stabilizer encoding and
syndrome correction attain the unrestricted optimum there.

\begingroup
\tcbset{commonstyle/.append style={unbreakable}}
\begin{boxcorollary}[Exact reliability at and above the critical rate]
\label{prop:reliability-matching}
Let $d$ be prime and $0<p<1/2$.  The critical rate is
\begin{equation}\label{eq:r_crit}
 r_{\rm crit}(p,d):=\frac{1-(d^2+1)p}{1+(d^2-1)p}.
\end{equation}
At every positive rate $r$ below capacity with $r\ge r_{\rm crit}(p,d)$,
all four reliability functions exist and take the same value.
Suppressing the common arguments $(r,p;d)$, we have
\begin{equation}\label{eq:exact-reliability-matching-range}
 E_{\STAB,\STAB}=E_{\STAB,\All}
 =E_{\All,\STAB}=E_{\All,\All}
 =\Dbin\!\left(\frac{1-r}{2}\middle\|p\right).
\end{equation}
If $p\ge1/(d^2+1)$, the critical rate is nonpositive and the formula
holds for every positive rate below capacity.
\end{boxcorollary}
\endgroup
\begin{proof}
Here $E_{\rm rc}$ equals the right side of
\eqref{eq:exact-reliability-matching-range} by
Proposition~\ref{prop:random-coding-closed-form}.  For every
$(\alpha,\beta)\in\{\STAB,\All\}^2$,
Theorem~\ref{thm:reliability-main} and the class ordering in
\eqref{eq:code-class-exponent-ordering} therefore give
\begin{equation}
 \begin{aligned}
 \Dbin((1-r)/2\|p)
 &\le\underline E_{\STAB,\STAB}
 \le\underline E_{\alpha,\beta}\\
 &\le\overline E_{\alpha,\beta}
 \le\overline E_{\All,\All}
 \le\Dbin((1-r)/2\|p).
 \end{aligned}
\end{equation}
Thus the lower and upper limits for all four classes equal
$\Dbin((1-r)/2\|p)$, so all four reliability functions exist and
have this common value.
If $p\ge1/(d^2+1)$, then $r_{\rm crit}(p,d)\le0$;
the matching region therefore includes every positive rate below capacity.
\end{proof}

Appendix~\ref{app:ostrev} compares two related tasks: physical Weyl
error guessing from the syndrome and erasure positions, extending
Ostrev's qubit formulation
\cite[Definitions~2--3 and Theorem~7]{ostrevCanonicalFormFinite2026a},
and Wilde's classical communication over an erasure channel without
feedback \cite[Proposition~8.7]{wildeFiniteblocklengthClassicalCommunication2026}.
At the corresponding parameters, both tasks share the quantum
reliability exponent in this rate region, though their finite-blocklength
errors can differ.  Wilde likewise uses random coding and expurgation
for achievability and derives the converse from feedback-assisted
performance \cite[Propositions~8.2 and~8.7 and Appendix~B]{wildeFiniteblocklengthClassicalCommunication2026}.

The proofs proceed in three steps.  Random-stabilizer coding gives the
first achievable lower bound, and logical-spectrum expurgation can improve
the bound at low rates.  A change of erasure probability then gives the
converse for arbitrary encoders and recoveries.  Together with the
closed form of the random-coding bound, these results establish
Corollary~\ref{prop:reliability-matching}.  Numerical illustrations
follow the core proofs.

\subsection{Achievability}
\subsubsection{Random-stabilizer coding}
We first bound the probability that a random stabilizer code has
nonzero logical ambiguity for a fixed erasure pattern, then average
over the number of erasures to obtain a reliability lower bound.

We specialize the random-stabilizer method of
\cite{hamadaLowerBoundsQuantum2002} to erasures.
\begin{boxlemma}[Random-stabilizer ambiguity bound]
\label{lem:random-stabilizer-ambiguity}
Let $\mu^{\mathrm S}_{n,k}$ denote the uniform distribution on the stabilizer
subspaces of $[[n,k]]_d$ codes, using the shorthand defined above:
\begin{equation}\label{eq:uniform-stabilizer-distribution}
 \mu^{\mathrm S}_{n,k}:=\operatorname{Unif}\{\Stab:\ [[n,k]]_d\},
 \qquad \Stab\sim\mu^{\mathrm S}_{n,k}.
\end{equation}
Thus the distribution is uniform over isotropic subspaces of $\Fd^{2n}$
with dimension $n-k$; the fixed local dimension $d$ is suppressed in its
notation.  Equivalently, choose a uniform ordered tuple
of $n-k$ linearly independent, pairwise symplectically orthogonal
vectors and take its span.  For every fixed $E\subseteq[n]$,
\begin{equation}
 \Pr_{\Stab\sim\mu^{\mathrm S}_{n,k}}\{\Amb(E)>0\}
 \le\min\{1,d^{2|E|-(n-k)}\}. \label{eq:random Stab uncorrectable error upper bound}
\end{equation}
\end{boxlemma}
\begin{proof}
Symplectic transformations are invertible linear maps on $\Fd^{2n}$
that preserve the symplectic form; they form the symplectic group.
The uniform ensemble is invariant under symplectic transformations,
and the symplectic group acts transitively on nonzero labels
\cite[Lemma~6 and Appendix~D]{hamadaLowerBoundsQuantum2002}; see also
\cite[Lemma~13]{ostrevCanonicalFormFinite2026a} for $d=2$.
Thus $\Pr\{v\in\Stab^\perp\setminus\Stab\}$ is the same for
all $v\ne0$.  Each stabilizer subspace has
$|\Stab^\perp\setminus\Stab|=d^{n+k}-d^{n-k}$.
Counting pairs $(\Stab,v)$ with
$v\in\Stab^\perp\setminus\Stab$ first by $\Stab$ and then by
$v\ne0$ therefore gives
\begin{equation}\label{eq:random-label}
 \Pr_{\Stab\sim\mu^{\mathrm S}_{n,k}}\{v\in\Stab^\perp\setminus\Stab\}
 =\frac{d^{n+k}-d^{n-k}}{d^{2n}-1}
 \le d^{-(n-k)}.
\end{equation}

Positive ambiguity occurs exactly when some
$v\in V_E\setminus\{0\}$ lies in $\Stab^\perp\setminus\Stab$.
The union bound
\cite[Ch.~1, Eq.~(1.11)]{gallagerDiscreteStochasticProcesses2011}
and $|V_E\setminus\{0\}|=d^{2|E|}-1$ yield
\begin{align}
 \Pr_{\Stab\sim\mu^{\mathrm S}_{n,k}}\{\Amb(E)>0\}
 &\le\sum_{v\in V_E\setminus\{0\}}
   \Pr_{\Stab\sim\mu^{\mathrm S}_{n,k}}\{v\in\Stab^\perp\setminus\Stab\}\notag\\
 &\le(d^{2|E|}-1)d^{-(n-k)}
 \le d^{2|E|-(n-k)}.
\end{align}
Combining this with the trivial upper bound one proves the claim.
\end{proof}

\begin{boxproposition}[Random-stabilizer reliability lower bound]
\label{thm:random-coding-reliability}
With $E_{\rm rc}$ defined in \eqref{eq:random-coding-exponent},
for $0\le r<1-2p$, it holds that
\begin{equation}\label{eq:random-coding-reliability}
\underline E_{\STAB,\STAB}(r,p;d)
 \ge E_{\rm rc}(r,p;d),
\end{equation}
\end{boxproposition}

\begin{proof}

Choose $k=\lceil nr\rceil$ and $\Stab\sim\mu^{\mathrm S}_{n,k}$
independently of $E_p$.  For each fixed $E$,
Lemma~\ref{lem:random-stabilizer-ambiguity} bounds the ensemble
probability of nonzero logical ambiguity by a function of $|E|$.
Average over $E_p$ and group the erasure patterns by their size $e$.
Since at least one code has an ambiguity probability no larger than the
ensemble average, the definition of $\eta_{n,\STAB,\STAB}^*(r,p;d)$ in
\eqref{eq: def eta_n^*(r,p)} gives
\begin{equation}\label{eq: achievable bound 1 eq1}
\begin{aligned}
 \eta_{n,\STAB,\STAB}^*(r,p;d)
 &\le\mathbb E_{\Stab\sim\mu^{\mathrm S}_{n,k}}
        \Pr_{E_p}\{a_{\Stab}(E_p)>0\}\\
 &\le\sum_{e=0}^n\binom ne p^e(1-p)^{n-e}
        \min\{1,d^{2e-(n-k)}\}.
\end{aligned}
\end{equation}
We next evaluate the exponential rate of this sum.  For $t=e/n$,
Lemma~\ref{lem:binary-type-tail-estimates}(a) in
Appendix~\ref{app:binomial-estimates} gives
\begin{equation}\label{eq: achievable bound 1 e bound}
 \binom ne p^e(1-p)^{n-e}\le e^{-n\Dbin(t\|p)}.
\end{equation}
Since $0\le k-nr<1$ and $x\mapsto x_+$ is 1-Lipschitz,
\begin{equation}\label{eq: achievable bound 1 uncorrectable bound}
 \min\{1,d^{2e-(n-k)}\}
 =e^{-n(1-k/n-2t)_+\ln d}
 \le d\,e^{-n(1-r-2t)_+\ln d}.
\end{equation}
There are $n+1$ types, and the minimum over them is at least
$E_{\rm rc}(r,p;d)$.  Therefore
\begin{equation}\label{eq:finite-random-coding-reliability}
 \eta_{n,\STAB,\STAB}^*(r,p;d)
 \le d(n+1)e^{-nE_{\rm rc}(r,p;d)}.
\end{equation}
Taking the lower exponential limit and applying
Lemma~\ref{lem:stabilizer-error-exponent-equivalence} proves the claim.  At $r=0$,
the stabilizer error is zero by definition, so the lower bound
remains true but does not identify a finite reliability function.
\end{proof}

The variational expression can be evaluated explicitly.  Set
\begin{equation}\label{eq:reliability-tau-critical}
 \tau_d(p):=\frac{d^2p}{1-p+d^2p},
 \qquad
 r_{\rm crit}(p,d)=1-2\tau_d(p),
\end{equation}
where the critical rate is defined in \eqref{eq:r_crit}.

\begin{boxproposition}[Closed form of the random-coding lower bound]
\label{prop:random-coding-closed-form}
For $0\le r<1-2p$,
\begin{equation}\label{eq:random-coding-closed-form}
 E_{\rm rc}(r,p;d)
 =\begin{cases}
 (1-r)\ln d-\ln(1-p+pd^2),
 &r\le r_{\rm crit}(p,d),\\[1ex]
 \Dbin\!\left(\dfrac{1-r}{2}\middle\|p\right),
 &r\ge r_{\rm crit}(p,d).
 \end{cases}
\end{equation}
If $r_{\rm crit}(p,d)<0$, only the second branch occurs at nonnegative
rates.
\end{boxproposition}

\begin{proof}
Put $a=(1-r)/2>p$ and let
$f(t)=\Dbin(t\|p)+(1-r-2t)_+\ln d$.  On $[a,1]$, $f(t)=\Dbin(t\|p)$,
whose minimum is $\Dbin(a\|p)$.  On $[0,a]$,
\begin{equation}
 f'(t)=\ln\frac{t(1-p)}{p(1-t)}-2\ln d,
\end{equation}
and $f''(t)=1/[t(1-t)]>0$ for $0<t<a$.
Thus the unique unconstrained minimizer on the linear-penalty branch
is $t=\tau_d(p)$.  Convexity shows
that the global minimizer is $\tau_d(p)$ when
$\tau_d(p)\le a$, equivalently $r\le r_{\rm crit}(p,d)$, and is $a$
otherwise.  With $Z=1-p+pd^2$,
\begin{equation}
 \frac{\tau_d(p)}p=\frac{d^2}{Z},\qquad
 \frac{1-\tau_d(p)}{1-p}=\frac1Z,
\end{equation}
so substitution into \eqref{eq:binary-relative-entropy} gives
\begin{equation}
 \Dbin(\tau_d(p)\|p)=2\tau_d(p)\ln d-\ln Z,
\end{equation}
which gives the first branch of \eqref{eq:random-coding-closed-form}; the
second is the value at $a$.
\end{proof}
\subsubsection{Logical-spectrum expurgation at low rates}
\label{subsubsec:logical-spectrum-expurgation}

We continue with the ensemble $\mu^{\mathrm S}_{n,k}$ from
\eqref{eq:uniform-stabilizer-distribution} and the error quantities
defined above.  Following the low-rate spectrum methods of
\cite{bargLowrateBoundReliability2002,ashikhminFidelityLowerBounds2014},
expurgation removes codes with low-weight nontrivial
logical labels before applying the ensemble average.
For a stabilizer subspace $\Stab:\ [[n,k]]_d$, define
\begin{equation}\label{eq:logical-weight-enumerator}
 L_w(\Stab):=\bigl|\{v\in\Stab^\perp\setminus\Stab:
                          \operatorname{wt}(v)=w\}\bigr|,
 \qquad 1\le w\le n.
\end{equation}
Here $\operatorname{wt}(v)$ is the number of physical qudits $i$ for
which the Weyl coordinate pair $(x_i,z_i)$ is nonzero.
Thus $L_w$ counts physical representatives with nontrivial logical
action; different representatives of the same logical coset are counted
separately.

A union bound over nontrivial logical labels gives
\begin{equation}\label{eq:logical-spectrum-union-bound}
 \Pr_{E_p}\{a_{\Stab}(E_p)>0\}\le\sum_{w=1}^n L_w(\Stab)p^w.
\end{equation}
The event $a_{\Stab}(E)>0$ means that at least one label counted by
the spectrum has its entire support in $E$.  The factor $p^w$ is the
probability that a weight-$w$ label has its support erased; coordinates
outside its support are unrestricted.

For $\Stab\sim\mu^{\mathrm S}_{n,k}$ and any fixed nonzero Weyl label
$v$, let $\pi_{n,k}$ be the probability that $v$ has nontrivial logical
action, namely $v\in\Stab^\perp\setminus\Stab$.  By \eqref{eq:random-label},
\begin{equation}\label{eq:logical-label-incidence-probability}
\begin{aligned}
 \pi_{n,k}
 &:=\Pr_{\Stab\sim\mu^{\mathrm S}_{n,k}}
       \{v\in\Stab^\perp\setminus\Stab\}\\
 &=\frac{d^{n+k}-d^{n-k}}{d^{2n}-1}\le d^{-(n-k)}.
\end{aligned}
\end{equation}
For $1\le w_0\le n$, let $\beta_{n,k,w_0}$ be the expected number of
nontrivial logical labels of weight less than $w_0$, counting physical
representatives as in $L_w$:
\begin{equation}\label{eq:expurgation-beta}
 \beta_{n,k,w_0}
 :=\mathbb E_{\Stab\sim\mu^{\mathrm S}_{n,k}}
       \!\left[\sum_{w=1}^{w_0-1}L_w(\Stab)\right].
\end{equation}

To evaluate this expectation, first count all weight-$w$ Weyl labels.
Each physical position $i$ carries a coordinate pair
$(x_i,z_i)\in\mathbb F_d^2$, and a label has weight $w$ exactly when
$w$ of these pairs are nonzero.  Choosing those $w$ positions gives
$\binom nw$ possible supports.  At each chosen position, the pair can
be any of the $d^2-1$ elements of
$\mathbb F_d^2\setminus\{(0,0)\}$; the choices at different positions
are independent, giving $(d^2-1)^w$ assignments on each support.
All other pairs must be $(0,0)$.  Each label determines its support
and these assignments uniquely, so there are exactly
$\binom nw(d^2-1)^w$ weight-$w$ Weyl labels.  This counts labels,
without an additional factor for an overall operator phase.

Each such label has logical incidence probability $\pi_{n,k}$ by
\eqref{eq:logical-label-incidence-probability}.
Linearity of expectation gives
\begin{equation}\label{eq:average-logical-weight-enumerator}
 \mathbb E_{\Stab\sim\mu^{\mathrm S}_{n,k}}L_w(\Stab)
 =\pi_{n,k}\binom nw(d^2-1)^w.
\end{equation}
Summing over $1\le w<w_0$ and using \eqref{eq:expurgation-beta}
therefore yields
\begin{equation}\label{eq:expurgation-beta-evaluation}
 \beta_{n,k,w_0}
 =\pi_{n,k}\sum_{w=1}^{w_0-1}\binom nw(d^2-1)^w.
\end{equation}

The following lemma specializes the conditional-spectrum method of
\cite[Theorem~6]{ashikhminFidelityLowerBounds2014} to erasures and
general local dimension $d$.  We include the proof to make this
adaptation explicit.

\begin{boxlemma}[Finite-blocklength logical-spectrum expurgation]
\label{prop:finite-logical-expurgation}
If $1\le w_0\le n$ and $\beta_{n,k,w_0}<1$, there is a stabilizer subspace
$\Stab:\ [[n,k]]_d$ with $L_w(\Stab)=0$ for $w<w_0$ satisfying
\begin{equation}\label{eq:finite-logical-expurgation}
  \Pr_{E_p}\{a_{\Stab}(E_p)>0\}
 \le\frac{\pi_{n,k}}{1-\beta_{n,k,w_0}}
       \sum_{w=w_0}^n\binom nw(d^2-1)^wp^w.
\end{equation}
\end{boxlemma}

\begin{proof}
Let $\mathcal A$ be the event that $L_w(\Stab)=0$ for every $w<w_0$.
The sum $\sum_{w=1}^{w_0-1}L_w(\Stab)$ is nonnegative, has mean
$\beta_{n,k,w_0}$, and is at least one on $\mathcal A^c$.
Markov's inequality therefore gives
\begin{equation}
 \Pr(\mathcal A)\ge1-\beta_{n,k,w_0}>0.
\end{equation}
Since the erasure-weighted sum is nonnegative, the definition of
conditional expectation and \eqref{eq:average-logical-weight-enumerator}
give
\begin{align}
 \mathbb E\!\left[\sum_{w=w_0}^n L_w(\Stab)p^w
                  \,\middle|\,\mathcal A\right]
 &\le\frac{\sum_{w=w_0}^n p^w\mathbb E L_w(\Stab)}
              {\Pr(\mathcal A)}\notag\\
 &\le\frac{\pi_{n,k}}{1-\beta_{n,k,w_0}}
       \sum_{w=w_0}^n\binom nw(d^2-1)^wp^w.
\end{align}
Some code in $\mathcal A$ has a weighted sum no larger than this
conditional average.  For that code, the terms with $w<w_0$ vanish,
so \eqref{eq:logical-spectrum-union-bound} proves
\eqref{eq:finite-logical-expurgation}.
\end{proof}

We use $H_{d^2}$, $\delta_{\rm GV}(r;d)$, and $E_{\rm expur}(r,p;d)$
as defined in \eqref{eq:q2-ary-entropy}--\eqref{eq:logical-gv-distance}
and \eqref{eq:logical-expurgated-spectral}.

\begin{boxproposition}[Expurgated reliability lower bound]
\label{thm:logical-expurgated-reliability}
For $0<r<1-2p$,
\begin{equation}\label{eq:logical-expurgated-reliability}
 \underline E_{\STAB,\STAB}(r,p;d)
 \ge E_{\rm expur}(r,p;d).
\end{equation}
\end{boxproposition}

\begin{proof}
Take $k=\lceil nr\rceil$ and put $w_0=\lceil n\delta\rceil$, initially
with $0<\delta<1-d^{-2}$.  Whenever $\beta_{n,k,w_0}<1$,
Lemma~\ref{prop:finite-logical-expurgation} gives a code at this $k$.
By \eqref{eq: def eta_n^*(r,p)}, its uncorrectable-pattern probability
upper bounds the minimum, so
\begin{equation}
 \eta_{n,\STAB,\STAB}^*(r,p;d)
 \le\frac{\pi_{n,k}}{1-\beta_{n,k,w_0}}
       \sum_{w=w_0}^n\binom nw(d^2-1)^wp^w.
\end{equation}

Each summand is the number of weight-$w$ Weyl labels times the
probability that their support is erased.  For $\omega=w/n$,
Lemma~\ref{lem:binary-type-tail-estimates} gives
$\binom nw\le e^{nh(\omega)}$, including the endpoints by continuity.
Thus
\begin{equation}
\begin{aligned}
 a_w&:=\binom nw(d^2-1)^wp^w\\
 &\le\exp[n(h(\omega)+\omega\ln(d^2-1)+\omega\ln p)]\\
 &=\exp[n(H_{d^2}(\omega)+\omega\ln p)].
\end{aligned}
\end{equation}

To verify the screening condition, note that $\beta_{n,k,w_0}$ counts
labels without the erasure factor $p^w$.  If $w<w_0$, then
$w/n<\delta<1-d^{-2}$, where $H_{d^2}$ is increasing.
Hence $\binom nw(d^2-1)^w\le e^{nH_{d^2}(\delta)}$.  There are
$w_0-1\le n$ terms, so $\pi_{n,k}\le d^{-(n-k)}$ yields
\begin{align}
 \beta_{n,k,w_0}
 &\le(w_0-1)\exp[-(n-k)\ln d+nH_{d^2}(\delta)]\notag\\
 &\le n\exp[-(n-k)\ln d+nH_{d^2}(\delta)]\notag\\
 &\le dn\exp[-n((1-r)\ln d-H_{d^2}(\delta))].
\end{align}
The last step uses $k<nr+1$.  This bound tends to zero if
$H_{d^2}(\delta)<(1-r)\ln d$.  By \eqref{eq:logical-gv-distance} and
the strict monotonicity of $H_{d^2}$ on the interval under consideration,
\begin{equation}
 H_{d^2}(\delta)<(1-r)\ln d
 \quad\Longleftrightarrow\quad\delta<\delta_{\rm GV}(r;d).
\end{equation}
We therefore fix any $0<\delta<\delta_{\rm GV}(r;d)$ for the remainder
of the proof.  Then $\beta_{n,k,w_0}\to0$ exponentially, so
$\beta_{n,k,w_0}<1$ and the lemma applies for all sufficiently large $n$.

The remaining sum has $n-w_0+1\le n$ terms.  Since $w\ge w_0$ implies
$w/n\ge\delta$, its largest term gives
\begin{equation}
\begin{aligned}
 \sum_{w=w_0}^n a_w
 &\le(n-w_0+1)\max_{w_0\le w\le n}a_w\\
 &\le n\exp\left[n\max_{\delta\le\omega\le1}
       \{H_{d^2}(\omega)+\omega\ln p\}\right].
\end{aligned}
\end{equation}
Combining this with the lemma, the incidence bound, and
$\eta_{n,\STAB,\STAB}^*(r,p;d)\le1$ gives
\begin{equation}\label{eq:expurgated-type-sum}
 \eta_{n,\STAB,\STAB}^*
 \le\min\left\{1,\frac{n}{1-\beta_{n,k,w_0}}
 \exp\left[-(n-k)\ln d+n\max_{\delta\le\omega\le1}
 \{H_{d^2}(\omega)+\omega\ln p\}\right]\right\}.
\end{equation}
Here the numerator $n$ comes from the number of summands; the factor
$\pi_{n,k}$ contributes $-(n-k)\ln d$ to the exponential term.

Taking negative logarithms and dividing by $n$ yields
\begin{equation}
 -\frac1n\ln\eta_{n,\STAB,\STAB}^*(r,p;d)
 \ge\left[\begin{aligned}
 &(1-k/n)\ln d
   -\max_{\delta\le\omega\le1}
       \{H_{d^2}(\omega)+\omega\ln p\}\\
 &-\frac1n\ln\frac{n}{1-\beta_{n,k,w_0}}
 \end{aligned}\right]_+.
\end{equation}
The prefactor does not affect the limiting bound, since
\begin{equation}
 \frac1n\ln\frac{n}{1-\beta_{n,k,w_0}}
 =\frac{\ln n}{n}-\frac{\ln(1-\beta_{n,k,w_0})}{n}\longrightarrow0.
\end{equation}
In general, it suffices that $\beta_{n,k,w_0}<1$ eventually and
$n^{-1}\ln(1/(1-\beta_{n,k,w_0}))\to0$; the convergence to zero proved
above guarantees both conditions.
For this fixed $\delta$, take the lower limit as $n\to\infty$.
Using $k/n\to r$ and Lemma~\ref{lem:stabilizer-error-exponent-equivalence},
we obtain
\begin{equation}
 \underline E_{\STAB,\STAB}(r,p;d)
 \ge\left[(1-r)\ln d-
       \max_{\delta\le\omega\le1}
       \{H_{d^2}(\omega)+\omega\ln p\}\right]_+.
\end{equation}
This holds for every $\delta<\delta_{\rm GV}(r;d)$.  By continuity,
letting $\delta\uparrow\delta_{\rm GV}(r;d)$ after the blocklength limit gives
\begin{equation}
 \underline E_{\STAB,\STAB}(r,p;d)
 \ge\left[(1-r)\ln d-
          \max_{\delta_{\rm GV}(r;d)\le\omega\le1}
             \{H_{d^2}(\omega)+\omega\ln p\}\right]_+
 =E_{\rm expur}(r,p;d).
\end{equation}
The last equality follows from \eqref{eq:logical-expurgated-spectral},
which proves the proposition.
\end{proof}

The logical-spectrum expurgated lower bound admits a piecewise
expression in terms of the GV distance.  Write
\begin{equation}\label{eq:expurgated-omega-p}
 \omega_p:=\frac{(d^2-1)p}{1+(d^2-1)p},
\end{equation}
and define the two rate thresholds
\begin{equation}\label{eq:lex-rate-thresholds}
 r_{\rm expur}(p,d):=1-\frac{H_{d^2}(\omega_p)}{\ln d},
 \qquad
 r_0(p,d):=1-\frac{\ln(1+(d^2-1)p)}{\ln d}.
\end{equation}

\begin{boxproposition}[Closed form of the logical-spectrum expurgated lower bound]
\label{prop:lex-closed-form}
For $0<p<1/2$ and $0\le r<1-2p$,
\begin{equation}\label{eq:lex-closed-form}
 E_{\rm expur}(r,p;d)
 =\begin{cases}
  \delta_{\rm GV}(r;d)\ln(1/p),
  &r\le r_{\rm expur}(p,d),\\[1ex]
  (1-r)\ln d-\ln(1-p+pd^2),
  &r_{\rm expur}(p,d)<r<r_0(p,d),\\[1ex]
  0,&r\ge r_0(p,d).
 \end{cases}
\end{equation}
The thresholds satisfy $r_{\rm expur}(p,d)<r_0(p,d)$, and the branch
expressions agree at their boundaries whenever these lie in the
stated rate range.  If $r_{\rm expur}(p,d)<0$, the first branch is absent
at nonnegative rates.  If $r_0(p,d)\le0$, equivalently
$p\ge1/(d+1)$, the lower bound is zero throughout the stated rate range.
\end{boxproposition}

\begin{proof}
For $0<\omega<1$,
\begin{equation}
 \frac{d}{d\omega}\bigl[H_{d^2}(\omega)+\omega\ln p\bigr]
 =\ln\frac{(d^2-1)p(1-\omega)}{\omega}.
\end{equation}
This derivative is positive for $\omega<\omega_p$ and negative for
$\omega>\omega_p$.  Thus the maximum over
$[\delta_{\rm GV}(r;d),1]$ is attained at $\delta_{\rm GV}(r;d)$
when $\delta_{\rm GV}(r;d)\ge\omega_p$, and at $\omega_p$ otherwise.
Since $p<1$, both $\omega_p$ and $\delta_{\rm GV}(r;d)$ lie in the
increasing interval $[0,1-d^{-2}]$ of $H_{d^2}$.  Therefore
\begin{align}
 \delta_{\rm GV}(r;d)\ge\omega_p
 &\quad\Longleftrightarrow\quad
 (1-r)\ln d\ge H_{d^2}(\omega_p)\notag\\
 &\quad\Longleftrightarrow\quad r\le r_{\rm expur}(p,d).
\end{align}
In this case, \eqref{eq:logical-gv-distance} gives
\begin{align}
 E_{\rm expur}(r,p;d)
 &=\bigl[(1-r)\ln d-H_{d^2}(\delta_{\rm GV}(r;d))
          -\delta_{\rm GV}(r;d)\ln p\bigr]_+\notag\\
 &=\delta_{\rm GV}(r;d)\ln(1/p).
\end{align}
For $r>r_{\rm expur}(p,d)$, direct substitution gives
\begin{equation}
 H_{d^2}(\omega_p)+\omega_p\ln p
 =\ln(1+(d^2-1)p),
\end{equation}
so the spectral formula reduces to
\begin{equation}
 E_{\rm expur}(r,p;d)
 =\bigl[(1-r)\ln d-\ln(1+(d^2-1)p)\bigr]_+.
\end{equation}
The expression inside the positive part is positive exactly when
$r<r_0(p,d)$, proving the last two branches.
Finally, the same substitution yields
\begin{equation}
 r_0(p,d)-r_{\rm expur}(p,d)
 =\frac{\omega_p\ln(1/p)}{\ln d}>0.
\end{equation}
At $r=r_{\rm expur}(p,d)$, the first two expressions both equal
$\omega_p\ln(1/p)$; at $r=r_0(p,d)$, the second expression is zero.
The condition $r_0(p,d)\le0$ is equivalent to
$1+(d^2-1)p\ge d$, or $p\ge1/(d+1)$.
\end{proof}
\subsection{Converse for arbitrary encoders and recoveries}
The converse is an upper bound on the reliability achievable by any
protocol.  It combines a coding converse at a noisier channel
with a classical change of measure: we compare the probabilities of
the same erasure patterns under two different erasure probabilities,
while holding the protocol fixed.
Fix a target rate $r$.  For each blocklength $n$, let
$(\cEnc_n,\cRec_n)$ denote the encoder and recovery map of an
unrestricted protocol as in the preceding section, with $k_n$ logical
qudits and $k_n/n\ge r$; the subscripts indicate dependence on $n$.
The conditional infidelity and its average over
erasure patterns are given by
\eqref{eq:conditional-entanglement-infidelity} and
\eqref{eq:conditional-error-decomposition}.
The error of this chosen protocol need not be the optimal error
$\varepsilon_{n,\All,\All}^*(r,p;d)$.  When changing the erasure
probability, keep $\cEnc_n$ and $\cRec_n$ fixed.  The
fixed-pattern channel $\cEra_E$ is independent of $p$, so only
the distribution of the erased set changes and the conditional
infidelities remain fixed.

\begin{boxlemma}[Bernoulli change of measure]
\label{lem:reliability-change-measure}
For a fixed protocol sequence as above, let $0<p<p'<1$.
If $\liminf_{n\to\infty}\varepsilon_{\mathrm{ent}}(\cEnc_n,\cRec_n,p')\ge c>0$, then
\begin{equation}\label{eq:reliability-change-measure}
 \limsup_{n\to\infty}-\frac1n\ln\varepsilon_{\mathrm{ent}}(\cEnc_n,\cRec_n,p)
 \le\Dbin(p'\|p).
\end{equation}
\end{boxlemma}
\begin{proof}
For fixed $\delta>0$, put
\begin{equation}
 T_{n,\delta}
 =\left\{E\subseteq[n]:\left|\frac{|E|}{n}-p'\right|\le\delta\right\}.
\end{equation}
Lemma~\ref{lem:binary-type-tail-estimates}(c), applied at $p'$, implies
$\Pr(E_{p'}\notin T_{n,\delta})\to0$.
The hypothesis gives $N_1$ such that, for all $n\ge N_1$,
$\varepsilon_{\mathrm{ent}}(\cEnc_n,\cRec_n,p')\ge3c/4$.
Since the tail probability tends to zero, there is
$N_2$ such that $\Pr(E_{p'}\notin T_{n,\delta})\le c/4$ for all
$n\ge N_2$.  Thus, for every $n\ge\max\{N_1,N_2\}$,
\begin{equation}
 \varepsilon_{\mathrm{ent}}(\cEnc_n,\cRec_n,p')\ge\frac{3c}{4},
 \qquad \Pr(E_{p'}\notin T_{n,\delta})\le\frac c4.
\end{equation}
For these $n$, the contribution outside $T_{n,\delta}$ satisfies
\begin{equation}
 0\le\sum_{E\notin T_{n,\delta}}\Pr(E_{p'}=E)\,
 \varepsilon_{\mathrm{ent}}(\cEnc_n,\cRec_n\mid E)
 \le\Pr(E_{p'}\notin T_{n,\delta})\le\frac c4,
\end{equation}
where the first two inequalities use $0\le \varepsilon_{\mathrm{ent}}(\cEnc_n,\cRec_n\mid E)\le1$.
Splitting the average error into the contributions inside and
outside $T_{n,\delta}$ therefore gives
\begin{align}
 &\sum_{E\in T_{n,\delta}}\Pr(E_{p'}=E)\,
 \varepsilon_{\mathrm{ent}}(\cEnc_n,\cRec_n\mid E)\notag\\
 =&\varepsilon_{\mathrm{ent}}(\cEnc_n,\cRec_n,p')
   -\sum_{E\notin T_{n,\delta}}\Pr(E_{p'}=E)\,
   \varepsilon_{\mathrm{ent}}(\cEnc_n,\cRec_n\mid E)\notag\\
 \ge&\varepsilon_{\mathrm{ent}}(\cEnc_n,\cRec_n,p')-\Pr(E_{p'}\notin T_{n,\delta})\notag\\
 \ge&\frac{3c}{4}-\frac c4=\frac c2.
\end{align}
For $t=|E|/n$, direct expansion gives
\begin{align}
 \frac1n\ln\frac{p^{|E|}(1-p)^{n-|E|}}
                       {(p')^{|E|}(1-p')^{n-|E|}}
 &=t\ln\frac p{p'}+(1-t)\ln\frac{1-p}{1-p'}\notag\\
 &=p'\ln\frac p{p'}+(1-p')\ln\frac{1-p}{1-p'}
   +(t-p')\left(\ln\frac p{p'}-\ln\frac{1-p}{1-p'}\right)\notag\\
 &=-\Dbin(p'\|p)
   -(t-p')\ln\frac{p'(1-p)}{p(1-p')}.
\end{align}
On $T_{n,\delta}$, the last term is bounded below by minus $\delta$
times the absolute logarithm, so
\begin{equation}
 \frac{p^{|E|}(1-p)^{n-|E|}}{(p')^{|E|}(1-p')^{n-|E|}}
 \ge\exp\left[-n\left(\Dbin(p'\|p)+
 \delta\left|\ln\frac{p'(1-p)}{p(1-p')}\right|\right)\right].
\end{equation}
Multiplying by the nonnegative values $\varepsilon_{\mathrm{ent}}(\cEnc_n,\cRec_n\mid E)$ and summing over
$T_{n,\delta}$ bounds its contribution to $\varepsilon_{\mathrm{ent}}(\cEnc_n,\cRec_n,p)$
from below.  All remaining terms are nonnegative, so
\begin{align}
 \varepsilon_{\mathrm{ent}}(\cEnc_n,\cRec_n,p)
 &\ge\sum_{E\in T_{n,\delta}}\Pr(E_p=E)\,
 \varepsilon_{\mathrm{ent}}(\cEnc_n,\cRec_n\mid E)\notag\\
 &\ge\exp\left[-n\left(\Dbin(p'\|p)+
 \delta\left|\ln\frac{p'(1-p)}{p(1-p')}\right|\right)\right]\notag\\
 &\quad{}\times\sum_{E\in T_{n,\delta}}\Pr(E_{p'}=E)\,
 \varepsilon_{\mathrm{ent}}(\cEnc_n,\cRec_n\mid E)\notag\\
 &\ge\frac c2\exp\left[-n\left(\Dbin(p'\|p)+
 \delta\left|\ln\frac{p'(1-p)}{p(1-p')}\right|\right)\right].
\end{align}
The last bound is positive.  Taking $-n^{-1}\ln$ reverses the
inequality and gives
\begin{equation}
 -\frac1n\ln\varepsilon_{\mathrm{ent}}(\cEnc_n,\cRec_n,p)
 \le\Dbin(p'\|p)+\delta\left|\ln\frac{p'(1-p)}{p(1-p')}\right|
 -\frac1n\ln\frac c2.
\end{equation}
Since $-n^{-1}\ln(c/2)\to0$, taking the limsup yields
\begin{equation}
 \limsup_{n\to\infty}-\frac1n\ln\varepsilon_{\mathrm{ent}}(\cEnc_n,\cRec_n,p)
 \le\Dbin(p'\|p)+\delta\left|\ln\frac{p'(1-p)}{p(1-p')}\right|.
\end{equation}
This holds for every fixed $\delta>0$.  Letting $\delta\downarrow0$
proves \eqref{eq:reliability-change-measure}.
\end{proof}

The following finite-blocklength bound ensures that the optimal
error is strictly positive at every positive rate.
\begin{boxlemma}[All-erased error lower bound]
\label{lem:all-erased-error-bound}
For every $n\ge1$, $0<p<1$, and $r>0$,
\begin{equation}\label{eq:all-code-error-positive}
 \varepsilon_{n,\All,\All}^*(r,p;d)
 \ge p^n(1-d^{-2nr})>0.
\end{equation}
\end{boxlemma}
\begin{proof}
Consider any unrestricted protocol with $k_n/n\ge r$, and write
$M_n=d^{k_n}\ge d^{nr}$ for its logical dimension.  The pattern
$E=[n]$ occurs with probability $p^n$.
Since the encoder is trace preserving and acts only on the sender's
system, the reference marginal remains $I_R/M_n$.  Conditioned on
$E=[n]$, the channel discards all physical systems and outputs
$|e\rangle\langle e|^{\otimes n}$, independently of the input.
Thus the joint state before recovery is
$(I_R/M_n)\otimes|e\rangle\langle e|^{\otimes n}$, and the recovered
state is
\begin{equation}
 \frac{I_R}{M_n}\otimes\sigma_n,
 \qquad
 \sigma_n:=\cRec_n\!\left(|e\rangle\langle e|^{\otimes n}\right),
 \qquad \tr\sigma_n=1.
\end{equation}
Expanding $|\Phi_{M_n}\rangle=M_n^{-1/2}
\sum_{j=1}^{M_n}|j\rangle_R|j\rangle_A$, its conditional
entanglement fidelity is
\begin{align}
 \langle\Phi_{M_n}|\left(\frac{I_R}{M_n}\otimes\sigma_n\right)
 |\Phi_{M_n}\rangle
 &=\frac1{M_n^2}\sum_{i,j=1}^{M_n}
   \langle i|I_R|j\rangle\langle i|\sigma_n|j\rangle\notag\\
 &=\frac1{M_n^2}\sum_{j=1}^{M_n}\langle j|\sigma_n|j\rangle
 =\frac1{M_n^2}.
\end{align}
Consequently, $\varepsilon_{\mathrm{ent}}(\cEnc_n,\cRec_n\mid[n])=1-M_n^{-2}$.  All other terms in
\eqref{eq:conditional-error-decomposition} are nonnegative, so
\begin{equation}
 \varepsilon_{\mathrm{ent}}(\cEnc_n,\cRec_n,p)\ge p^n \varepsilon_{\mathrm{ent}}(\cEnc_n,\cRec_n\mid[n])
 =p^n(1-M_n^{-2})\ge p^n(1-d^{-2nr}),
\end{equation}
where the last inequality uses $M_n\ge d^{nr}$.
Taking the infimum over all admissible protocols preserves this
uniform lower bound and proves \eqref{eq:all-code-error-positive}.
Strict positivity follows from $p,r>0$ and $d>1$.
\end{proof}

We use the pretty strong converse for degradable channels
\cite[Theorem~2]{morganPrettyStrongConverse2014}.
A complementary channel gives the environment output in an
isometric realization of a channel.  A channel is degradable if a
CPTP map applied to the receiver output reproduces that environment
output.  The pretty strong converse used here bounds fidelity away
from one at rates above capacity; it does not require error to tend
to one.  In that reference the error is purified distance, which
for a state $\rho$ and our pure target $|\Phi_M\rangle$ equals
$\sqrt{1-\langle\Phi_M|\rho|\Phi_M\rangle}$.

\begin{boxproposition}[Reliability upper bound for arbitrary codes]
\label{prop:reliability-converse}
For $0<p<1/2$ and $0<r<1-2p$,
\begin{equation}\label{eq:reliability-converse}
 \overline E_{\All,\All}(r,p;d)
 \le\Dbin\!\left(\frac{1-r}{2}\middle\|p\right).
\end{equation}
\end{boxproposition}
\begin{proof}
By Lemma~\ref{lem:all-erased-error-bound}, the optimal error is positive.
Hence, for each $n$, choose a protocol from the unrestricted encoder
class in \eqref{eq:all-code-optimal-error}, with $k_n$ logical qudits,
logical dimension $M_n=d^{k_n}$, and rate $k_n/n\ge r$, such that
\begin{equation}
 \varepsilon_{n,\All,\All}^*(r,p;d)
 \le\varepsilon_{\mathrm{ent}}(\cEnc_n,\cRec_n,p)
 \le2\varepsilon_{n,\All,\All}^*(r,p;d).
\end{equation}

We next obtain an asymptotic error lower bound at a noisier channel.
Fix $p'\in((1-r)/2,1/2)$; since $r<1-2p$, we have $p'>p$.
Evaluate $\varepsilon_{\mathrm{ent}}(\cEnc_n,\cRec_n,p')$ with the same protocol sequence,
keeping $M_n$, $\cEnc_n$, and $\cRec_n$ fixed. Each protocol is an entanglement-generation protocol in the sense
of \cite[Section~II]{morganPrettyStrongConverse2014}: the sender
prepares $(\id_R\otimes\cEnc_n)(\Phi_{M_n})$, retains $R$, and
sends the physical system through the channel; the receiver
applies $\cRec_n$.  Its recovered state and target $\Phi_{M_n}$
are exactly those defining $\varepsilon_{\mathrm{ent}}(\cEnc_n,\cRec_n,p')$.
In Morgan and Winter's convention, the purified-distance error
for this pure target is therefore $\sqrt{\varepsilon_{\mathrm{ent}}(\cEnc_n,\cRec_n,p')}$.

To apply their pretty strong converse, note that the erasure channel
at $p'<1/2$ is degradable
\cite[Section~IV]{morganPrettyStrongConverse2014}, with normalized
capacity $1-2p'<r$ by \eqref{eq:erasure-capacity}.
If $\varepsilon_{\mathrm{ent}}(\cEnc_n,\cRec_n,p')\le1/2-\delta$ on an infinite
subsequence for some fixed $\delta\in(0,1/2)$, the purified distance is at most
$\sqrt{1/2-\delta}<1/\sqrt2$.  The pretty strong converse for degradable channels
\cite[Theorem~2]{morganPrettyStrongConverse2014}, expressed in nats,
would then give
\begin{equation}
 nr\ln d\le\ln M_n
 \le n(1-2p')\ln d+O(\sqrt{n\ln n}).
\end{equation}
Here $O(b_n)$ denotes a term whose absolute value is bounded by a
fixed multiple of $b_n$ for all sufficiently large $n$.
Dividing by $n$ contradicts $r>1-2p'$ for sufficiently large $n$.
Hence $\liminf_{n\to\infty}\varepsilon_{\mathrm{ent}}(\cEnc_n,\cRec_n,p')\ge1/2$.

This verifies the hypothesis of Lemma~\ref{lem:reliability-change-measure}
with $c=1/2$.  Applying it to the same fixed protocol sequence gives
\begin{equation}
 \limsup_{n\to\infty}-\frac1n\ln\varepsilon_{\mathrm{ent}}(\cEnc_n,\cRec_n,p)
 \le\Dbin(p'\|p).
\end{equation}

To transfer this bound to the optimal error, use
$\varepsilon_{\mathrm{ent}}(\cEnc_n,\cRec_n,p)\le2\varepsilon_{n,\All,\All}^*(r,p;d)$ to obtain
\begin{equation}
 -\frac1n\ln\varepsilon_{n,\All,\All}^*(r,p;d)
 \le-\frac1n\ln\varepsilon_{\mathrm{ent}}(\cEnc_n,\cRec_n,p)+\frac{\ln2}{n}.
\end{equation}
Taking the limsup gives
$\overline E_{\All,\All}(r,p;d)\le\Dbin(p'\|p)$.
Finally, let $p'\downarrow(1-r)/2$ and use continuity of $\Dbin$
to obtain \eqref{eq:reliability-converse}.
\end{proof}
\label{subsubsec:reliability-matching}

For $r\ge r_{\rm crit}(p,d)$, Proposition~\ref{prop:random-coding-closed-form}
gives $E_{\rm rc}(r,p;d)=\Dbin((1-r)/2\|p)$, matching the converse
in Proposition~\ref{prop:reliability-converse}.  The resulting exact
reliability functions are stated in
Corollary~\ref{prop:reliability-matching}.

We now show that the lower and upper bounds in
\eqref{eq:reliability-main-bounds} are strictly separated at every
positive rate below the critical rate.  For $0<r<1-2p$ with
$r<r_{\rm crit}(p,d)$, put $a=(1-r)/2$.
The minimizer $\tau_d(p)$ in the proof of
Proposition~\ref{prop:random-coding-closed-form} lies strictly below $a$,
so strict convexity gives $E_{\rm rc}(r,p;d)<\Dbin(a\|p)$.
For the expurgated bound in Proposition~\ref{prop:lex-closed-form},
the linear branch equals $E_{\rm rc}$ and the zero branch is strictly
below $\Dbin(a\|p)$ because $a>p$.

It remains to compare $\delta\ln(1/p)$ with $\Dbin(a\|p)$, where
$\delta=\delta_{\rm GV}(r;d)$.  At fixed $r$, this branch occurs for
$p\le p_x:=\delta/[(1-\delta)(d^2-1)]$, since its condition is
$\omega_p\le\delta$.  The strict concavity of $H_{d^2}$ between
$(0,0)$ and $(1-d^{-2},2\ln d)$ and
$H_{d^2}(\delta)=2a\ln d$ give $\delta<(1-d^{-2})a$.
Thus $\tau_d(p_x)=\delta/(1-d^{-2})<a$, so $p_x$ is below the
critical curve and the two nonzero branches agree there.  In particular,
$\delta\ln(1/p_x)=E_{\rm rc}(r,p_x;d)<\Dbin(a\|p_x)$.
For $0<p\le p_x$,
\begin{equation}
 \frac{d}{dp}\bigl[\Dbin(a\|p)-\delta\ln(1/p)\bigr]
 =\frac{(1-\delta)p-(a-\delta)}{p(1-p)}
 \le\frac{\delta/(1-d^{-2})-a}{p(1-p)}<0.
\end{equation}
The gap is therefore positive throughout this branch, since it is
positive at $p_x$.  Hence the combined lower bound in
\eqref{eq:reliability-main-bounds} is strictly below the converse at
all positive rates below $r_{\rm crit}(p,d)$.
\subsection{Numerical illustrations}
\label{subsubsec:reliability-numerics}

We first compare the regions in which the bounds in
\eqref{eq:reliability-main-bounds} coincide for $d=2,5,13$.  We then
illustrate slices at fixed erasure probability and fixed rate for $d=2$.

Corollary~\ref{prop:reliability-matching} and the strict comparison
above show that within $0<p<1/2$ and $0<r<1-2p$, the combined lower
bound $\max\{E_{\rm rc},E_{\rm expur}\}$ equals the converse exactly
when $r\ge r_{\rm crit}(p,d)$.  Solving this condition for $p$ gives
$p\ge p_{\rm crit}(r,d)$, where
\begin{equation}\label{eq:erasure-matching-threshold}
 p_{\rm crit}(r,d):=\frac{1-r}{d^2+1+(d^2-1)r}.
\end{equation}

\begin{figure}[!htb]
 \centering
 \includegraphics[width=\linewidth]{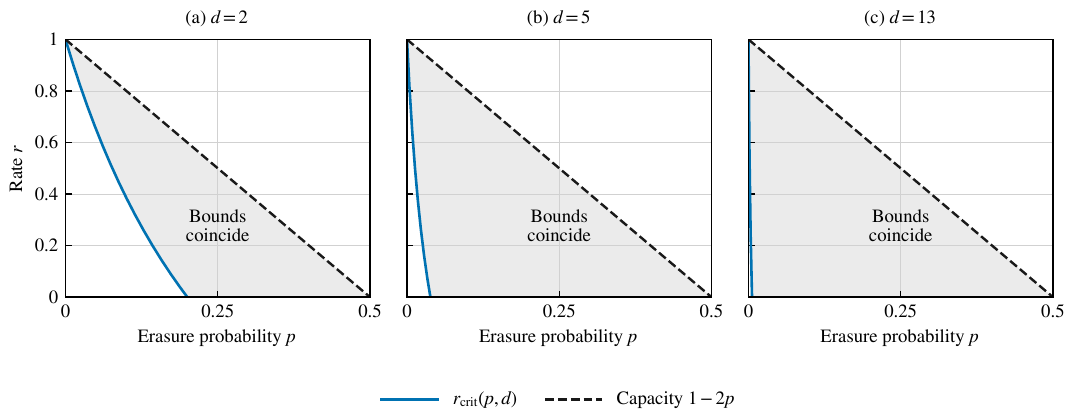}
 \caption{Coincident reliability bounds for (a) $d=2$, (b) $d=5$,
 and (c) $d=13$.  Shading marks $0<p<1/2$, $0<r<1-2p$, and
 $r\ge r_{\rm crit}(p,d)$.
 The vertical axis is the target normalized rate $r$; the corresponding
 target information rate is $r\ln d$ nats per channel use.
 The blue solid curve is the positive part of the critical-rate curve,
 equivalently $p=p_{\rm crit}(r,d)$ for $r>0$.  Where the critical rate
 is nonpositive, the lower boundary is $r=0$.  The black dashed line is
 the capacity boundary $r=1-2p$.
 The axes and capacity boundary are excluded from the stated region.}
 \label{fig:reliability-matching-region}
\end{figure}

Figure~\ref{fig:reliability-matching-region} shows the matching
region in the $(p,r)$ plane for $d=2,5,13$.  It lies between
$\max\{r_{\rm crit}(p,d),0\}$ and the capacity boundary $r=1-2p$,
including the critical-rate boundary at positive rates and excluding
the axes and capacity boundary.  The critical rate becomes
nonpositive at $p=1/(d^2+1)$, namely $1/5$, $1/26$, and $1/170$
for the three dimensions.  Above each threshold, the bounds coincide
throughout $0<r<1-2p$.  Increasing $d$ therefore enlarges the matching
region in these normalized coordinates.  In the shaded region,
\eqref{eq:exact-reliability-matching-range} gives the exact
reliability function.  Below the critical curve, the combined lower
bound is strictly smaller than the converse.

\begin{figure}[!htb]
 \centering
 \includegraphics[width=\linewidth]{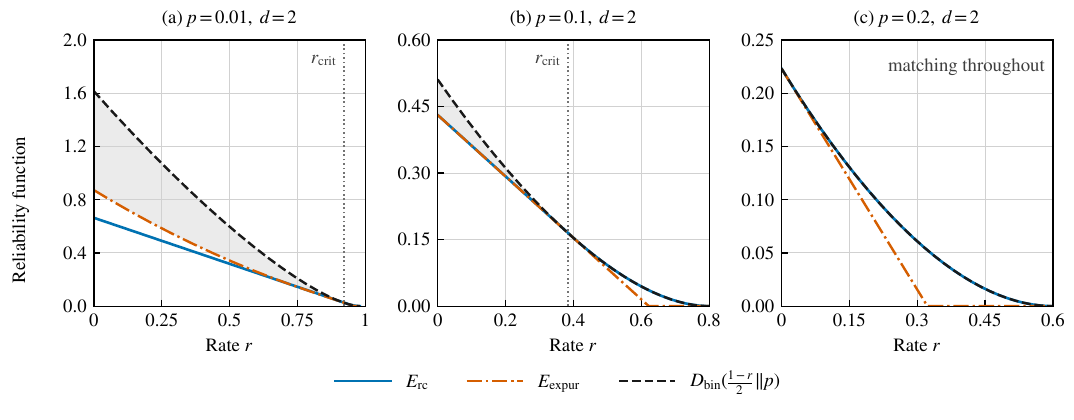}
 \caption{Reliability bounds for the qubit erasure channel as functions
 of $r$, at (a) $p=0.01$, (b) $p=0.1$, and (c) $p=0.2$.
 The blue solid, orange dash-dotted, and black dashed curves show
 $E_{\rm rc}$, $E_{\rm expur}$, and $\Dbin((1-r)/2\|p)$, respectively.
 Shading indicates the gap between the combined lower bound and the converse.
 The dotted lines in (a) and (b) mark $r_{\rm crit}(p,2)$ in
 \eqref{eq:r_crit}; in (c), this critical rate is zero and the
 random-coding bound is exact throughout.
 Curves are evaluated only for $0<r<1-2p$.}
 \label{fig:reliability-versus-rate}
\end{figure}

\begin{figure}[!htb]
 \centering
 \includegraphics[width=\linewidth]{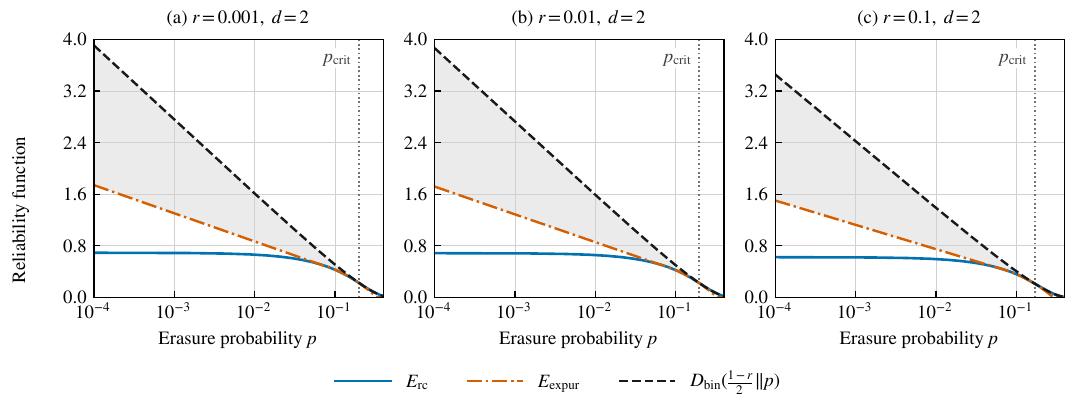}
 \caption{Reliability bounds for the qubit erasure channel as functions
 of $p$, at (a) $r=0.001$, (b) $r=0.01$, and (c) $r=0.1$.
 Curves and shading follow the conventions of
 Fig.~\ref{fig:reliability-versus-rate}.  The dotted lines labeled $p_{\rm crit}$ mark
 $p_{\rm crit}(r,d)$ in \eqref{eq:erasure-matching-threshold}, with $d=2$.
 The erasure-probability axis is logarithmic to resolve the improvement
 from expurgation at small $p$.  Only positive erasure probabilities
 are evaluated, with $0<p<(1-r)/2$.}
 \label{fig:reliability-versus-erasure}
\end{figure}

Figures~\ref{fig:reliability-versus-rate} and
\ref{fig:reliability-versus-erasure} evaluate the asymptotic bounds in
Propositions~\ref{prop:random-coding-closed-form} and
\ref{prop:lex-closed-form} in nats.  Shading shows the gap between
the combined lower bound and the converse.  Figure~\ref{fig:reliability-versus-rate}
takes vertical slices of Fig.~\ref{fig:reliability-matching-region}(a).
The critical rates at $p=0.01$, $0.1$, and $0.2$ are approximately
$0.9223$, $0.3846$, and $0$, respectively.  Expurgation improves
the lower bound at low rates for $p=0.01$.

Figure~\ref{fig:reliability-versus-erasure} takes horizontal slices of
Fig.~\ref{fig:reliability-matching-region}(a).  For each plotted rate,
expurgation improves the lower bound at small $p$.  The gap closes at
$p_{\rm crit}(r,2)=(1-r)/(5+3r)$, which approaches $1/5$ as $r$ tends
to zero.

\section{Conclusion}\label{sec:conclusion}

We studied how reliably quantum information can be transmitted through
a quantum erasure channel without assistance.  Each transmitted system
is independently erased with probability $p$, and the receiver learns
which positions were lost.  We measure error by the entanglement
infidelity: one minus the overlap between a maximally entangled target
state and the state obtained after transmission and recovery.  At a
fixed rate, the reliability function describes the optimal exponential
decay rate of this error as the number of channel uses grows.

For systems of prime dimension $d$, we establish lower and upper
bounds on the reliability function.  The lower bounds use randomly
chosen stabilizer codes, which encode into common eigenspaces of commuting
generalized Pauli operators, and expurgation, which discards codes containing
low-weight logical errors.  Measuring the stabilizer operators produces a
syndrome that guides error correction.  The upper bound applies to arbitrary quantum
encoders and recovery maps; it follows by comparing erasure patterns at
two erasure probabilities and invoking an error bound above capacity.
The two sides match at sufficiently high rates below capacity.  In that
region, the reliability function for stabilizer coding with syndrome
correction equals the reliability function for arbitrary quantum encoders
and recovery maps.

More precisely, for erasure probability $0<p<1/2$, a normalized rate $r$ is the number of
logical $d$-dimensional systems transmitted per channel use, and
capacity is $1-2p$ in these units.  The matching region consists of
positive rates below capacity and at least the critical rate
$r_{\rm crit}=(1-(d^2+1)p)/(1+(d^2-1)p)$.  In this region, the exact
reliability is the binary relative entropy, in natural logarithms, of a
Bernoulli distribution with parameter $(1-r)/2$ relative to one with
parameter $p$.  This quantity is the exponential cost of observing an
erased fraction near $(1-r)/2$ when each position is erased with
probability $p$.
When $p\ge1/(d^2+1)$, the region includes every positive rate below
capacity.

When $r_{\rm crit}>0$, the bounds established here do not determine the
exact reliability functions for positive rates below the critical rate.
Expurgation can improve the random-coding lower bound there, and the
combined lower bound remains strictly below the converse.  Improving either
bound would narrow this gap and could identify additional rates
at which the optimal error exponent is known.

\bigskip
\paragraph{Acknowledgements.}
This work is supported in part by the National Natural Science Foundation of China (Grant No. 92470113 and No. 12404569), the Shenzhen Science and Technology Program (Grant No. QNXMB20250701091826036 and No. JCYJ20240813113519025), the Shenzhen Fundamental Research Program (Grant No. JCYJ20241202124023031), the General R\&D Projects of 1+1+1 CUHK-CUHK(SZ)-GDST Joint Collaboration Fund (Grant No. GRDP2025-022), the Guangdong Provincial Quantum Science Strategic Initiative (Grant No. GDZX2603008 and No. GDZX2503001), and the University Development Fund (Grant No. UDF01003565).
OpenAI GPT-6 Astra was used to explore proof strategies and support the writing of the manuscript. We
have verified AI-assisted material and take full responsibility for the content of this work.

\begingroup
\bibliographystyle{alpha_abbrv}
\bibliography{ref}
\endgroup

\appendix
\section{Binomial estimates}\label{app:binomial-estimates}

We collect the probability estimates for the achievability and converse
analyses, together with the local expansion of binary relative entropy.
The notation $X_n\sim\operatorname{Binomial}(n,p)$ means that $X_n$
is the sum of $n$ independent random variables, each equal to one
with probability $p$ and zero otherwise.  Its probability mass at $e$ is
$\binom ne p^e(1-p)^{n-e}$, where $\binom ne$ counts the size-$e$
subsets of $[n]$.  Convergence in probability to $p$ means that
$\Pr\{|X_n/n-p|>\delta\}\to0$ for every $\delta>0$.

\begin{lemma}[Binary type and tail estimates]
\label{lem:binary-type-tail-estimates}
Fix $p\in(0,1)$ and let $X_n\sim\operatorname{Binomial}(n,p)$.
The integer $e$ below denotes the number of erasures; $\exp$ denotes
the natural exponential function, and $\Dbin$ is measured in nats.
The estimates below use the binomial coefficient bounds in
\cite[Lemma~9.2]{mitzenmacherProbabilityComputing2005} and
Chebyshev's inequality in \cite[Theorem~3.6]{mitzenmacherProbabilityComputing2005}.
\begin{enumerate}[label=(\alph*),leftmargin=2em]
 \item For every $e\in\{0,\ldots,n\}$ and $t=e/n$,
 \begin{equation}
  \frac{1}{n+1}\exp[-n\Dbin(t\|p)]
  \leq\Pr\{X_n=e\}
  \leq \exp[-n\Dbin(t\|p)].
  \label{eq:binary-type-point-bounds}
 \end{equation}
 \item If $a_n\to a$ with $p<a<1$, then either strict or non-strict
 upper tails have exponent
 \begin{equation}
  \lim_{n\to\infty}-\frac1n\ln
  \Pr\!\left\{\frac{X_n}{n}>a_n\right\}
  =\lim_{n\to\infty}-\frac1n\ln
  \Pr\!\left\{\frac{X_n}{n}\geq a_n\right\}
  =\Dbin(a\|p),
  \label{eq:binary-moving-tail-ldp}
 \end{equation}
 whenever the displayed events are nonempty for all sufficiently large
 $n$.
 \item $X_n/n\to p$ in probability.
 \item As $x\to0$,
 \begin{equation}
  \Dbin(p+x\|p)
  =\frac{x^2}{2p(1-p)}+O(x^3).
  \label{eq:binary-relative-entropy-local}
 \end{equation}
\end{enumerate}
\end{lemma}
\begin{proof}
By \cite[Lemma~9.2]{mitzenmacherProbabilityComputing2005}, expressed
in nats,
\begin{equation}
 \frac{1}{n+1}e^{nh(t)}
 \leq\binom ne\leq e^{nh(t)},
 \qquad
 h(t):=-t\ln t-(1-t)\ln(1-t),
\end{equation}
where $h(0)=h(1)=0$.  Multiplying by $p^e(1-p)^{n-e}$ and using
$h(t)+t\ln p+(1-t)\ln(1-p)=-\Dbin(t\|p)$ proves part~(a).

For part~(b), let $e_n$ be the smallest integer satisfying
$e_n/n>a_n$ (or $e_n/n\geq a_n$ for the non-strict tail).  Then
$e_n/n\to a$.  The lower bound follows by retaining the single term
$\Pr\{X_n=e_n\}$ and applying part~(a).  For the upper bound, sum the
upper estimate in part~(a) over at most $n+1$ types.  Since
$\partial_t\Dbin(t\|p)=\ln[t(1-p)/(p(1-t))]>0$ on $(p,1)$,
the smallest exponent in
the upper tail converges to $\Dbin(a\|p)$.  The polynomial factor $n+1$
does not affect the normalized logarithm.  For part~(c),
$\mathbb E(X_n/n)=p$ and $\operatorname{Var}(X_n/n)=p(1-p)/n$.
Chebyshev's inequality
\cite[Theorem~3.6]{mitzenmacherProbabilityComputing2005} gives, for every
$\delta>0$,
\begin{equation}
 \Pr\{|X_n/n-p|\ge\delta\}\le\frac{p(1-p)}{n\delta^2}\longrightarrow0.
\end{equation}
For part~(d), the value and first derivative at $t=p$
are zero, while
\begin{equation}
 \left.\partial_t^2\Dbin(t\|p)\right|_{t=p}
 =\left.\left(\frac1t+\frac1{1-t}\right)\right|_{t=p}
 =\frac1{p(1-p)}.
\end{equation}
The third derivative is bounded on a sufficiently small closed
neighborhood of $p$, so Taylor's formula gives the stated remainder.

\end{proof}
\section{Comparison with other coding tasks}\label{subsec:other-coding-tasks}
\label{app:ostrev}
The two auxiliary tasks below have different success criteria from
entanglement transmission.  We compare them with the four encoder--recovery
classes $(\STAB,\STAB)$, $(\STAB,\All)$, $(\All,\STAB)$, and $(\All,\All)$.
Their definitions use $0\le r\le1$;
the common reliability bounds are stated for $0<p<1/2$ and
$0<r<1-2p$.

\subsection{Physical-error guessing}
Ostrev studies physical-error guessing and logical-coset decoding for
qubit stabilizer codes and derives finite-blocklength guessing bounds
\cite[Definitions~2--3 and Theorem~7]{ostrevCanonicalFormFinite2026a}.
We compare these tasks under erasures and establish guessing
reliability bounds for every prime $d$ using the methods of the main text.

\paragraph{Task definition and error exponents.}

\label{subsec:guessing}
For a fixed stabilizer code, this task reconstructs the entire Weyl
label $U$ from the erased set $E$ and syndrome $\sigma_{\Stab}(U)$.
Let $g$ be a classical guessing rule that, for each $E$ and attainable
syndrome $t\in\sigma_{\Stab}(V_E)$, chooses a label $g(E,t)\in V_E$
with $\sigma_{\Stab}(g(E,t))=t$.
Define its minimum probability of error by
\begin{equation}\label{eq:guess-fixed-code-error}
 \varepsilon_{\mathrm{guess}}^*(\Stab,p)
 :=\inf_g
   \Pr_{E_p,U}\{U\ne g(E_p,\sigma_{\Stab}(U))\},
\end{equation}
where the infimum is over the guessing rules just described.  Here $g$
produces a classical guess, while quantum recovery is described by a
recovery map.
Theorem~\ref{thm:fixed-encoder-optimal-guessing} below evaluates this minimum.
Minimize over stabilizer codes at rate at least $r$:
\begin{equation}\label{eq:guess-finite-errors}
 \varepsilon_{n,\mathrm{guess}}^*(r,p;d)
 :=\min_{\substack{\Stab:\ [[n,k]]_d\\ k/n\ge r}}
      \varepsilon_{\mathrm{guess}}^*(\Stab,p).
\end{equation}
For this task, define the lower and upper error exponents by
\begin{equation}\label{eq:guess-reliability-exponents}
\begin{aligned}
 \underline E_{\mathrm{guess}}(r,p;d)
 &:=\liminf_{n\to\infty}-\frac1n\ln\varepsilon_{n,\mathrm{guess}}^*(r,p;d),\\
 \overline E_{\mathrm{guess}}(r,p;d)
 &:=\limsup_{n\to\infty}-\frac1n\ln\varepsilon_{n,\mathrm{guess}}^*(r,p;d).
\end{aligned}
\end{equation}
We use $-\ln0=+\infty$ and denote the common value by
$E_{\mathrm{guess}}(r,p;d)$ only when these limits coincide.

Correctly guessing the physical label also identifies its logical
class.  For each code space, choose a Clifford stabilizer encoder and use
syndrome correction; its logical error is no greater than the physical
guessing error.  Optimizing over stabilizer encoders and recoveries gives
\begin{equation}\label{eq:guess-logical-ordering}
 \varepsilon_{n,\STAB,\STAB}^*(r,p;d)
 \le\varepsilon_{n,\mathrm{guess}}^*(r,p;d).
\end{equation}
The prime-$d$ guessing bounds are proved in the reliability-bounds
paragraph below.  Ostrev's qubit finite-blocklength
bounds are developed in \cite{ostrevCanonicalFormFinite2026a}.

Taking normalized negative logarithms and lower or upper limits gives
the exponent comparisons
\begin{equation}
 \underline E_{\mathrm{guess}}(r,p;d)
 \le\underline E_{\STAB,\STAB}(r,p;d),\qquad
 \overline E_{\mathrm{guess}}(r,p;d)
 \le\overline E_{\STAB,\STAB}(r,p;d).
\end{equation}
The comparisons with the other quantum classes follow from
\eqref{eq:code-class-exponent-ordering}.

\paragraph{Optimal guessing and comparison with logical recovery.}

Physical-error guessing reconstructs the entire Weyl label $U$ from
the revealed erased set $E$ and syndrome $\sigma_{\Stab}(U)$, as in
\eqref{eq:guess-fixed-code-error}.  Logical recovery only identifies
its action modulo stabilizer labels.  With $a_{\Stab}(E)$ as in
\eqref{eq:erasure-logical-ambiguity}, write
$b_{\Stab}(E):=\dim_{\Fd}(\Stab\cap V_E)$.
The following theorem gives the physical-error-guessing analogue of
Theorem~\ref{prop:fixed-encoder-optimal-recovery}.

\begin{theorem}[Optimal physical-error guessing for a fixed stabilizer encoder]
\label{thm:fixed-encoder-optimal-guessing}
Let $d$ be prime and fix an $[[n,k]]_d$ stabilizer code $\Stab$.
Given the erased set $E$ and the syndrome $\sigma_{\Stab}(U)$ of a
uniform label $U\in V_E$, every guessing rule $g$ defined above attains
the minimum physical-error-guessing probability of error.
Consequently, at erasure probability $p$,
\begin{equation}\label{eq:fixed-encoder-optimal-guessing-error}
 \varepsilon_{\mathrm{guess}}^*(\Stab,p)
 =\mathbb E_{E_p}\!\left[
   1-d^{-(a_{\Stab}(E_p)+b_{\Stab}(E_p))}\right].
\end{equation}
\end{theorem}
\begin{proof}
Fix $E$ and an attainable syndrome $t$.  Choose $u_t\in V_E$ with
$\sigma_{\Stab}(u_t)=t$.  The restricted syndrome map is linear,
with kernel $\Stab^\perp\cap V_E$, so its fiber at $t$ is
\begin{equation}
 \{u\in V_E:\sigma_{\Stab}(u)=t\}
 =u_t+(\Stab^\perp\cap V_E).
\end{equation}
The subspace $\Stab\cap V_E$ of this kernel has dimension
$b_{\Stab}(E)$, and the quotient has dimension $a_{\Stab}(E)$ by
\eqref{eq:erasure-logical-ambiguity}.  The dimension formula for a
quotient therefore shows that the kernel has dimension
$a_{\Stab}(E)+b_{\Stab}(E)$.  Its affine translate contains exactly
$d^{a_{\Stab}(E)+b_{\Stab}(E)}$ labels.

Since $U$ is uniform on $V_E$, its posterior given $E$ and $t$ is
uniform on this fiber.  The rule $g$ selects one label in the
fiber and thus succeeds with probability
$d^{-(a_{\Stab}(E)+b_{\Stab}(E))}$.  No deterministic guess can do
better, and a randomized guess is a mixture of deterministic guesses,
so it cannot improve this value.  This success probability does not
depend on $t$, and averaging over the syndrome preserves it.

The decoder is given $E_p$ and can choose a section for every pattern.
These sections attain all conditional minima simultaneously.  Taking
complements and averaging over $E_p$ proves
\eqref{eq:fixed-encoder-optimal-guessing-error}.
\end{proof}

Logical recovery succeeds on a coset of $\Stab\cap V_E$ containing
$d^{b_{\Stab}(E)}$ labels.  Comparing
\eqref{eq:fixed-encoder-optimal-guessing-error} with
Theorem~\ref{prop:fixed-encoder-optimal-recovery} gives
\begin{equation}\label{eq:guess-logical-difference}
 \varepsilon_{\mathrm{guess}}^*(\Stab,p)
 -\inf_{\cRec\in\mathrm{CPTP}(\mathcal H_B,\mathcal H_A)}
 \varepsilon_{\mathrm{ent}}(\cEnc_{\Stab},\cRec,p)
 =\mathbb E_{E_p}\!\left[
 d^{-a_{\Stab}(E_p)}(1-d^{-b_{\Stab}(E_p)})\right].
\end{equation}
Thus stabilizer labels supported on the erased set can obstruct an
exact physical guess while acting trivially on the logical state.
The following example shows that the two errors can differ even
after optimizing over codes at fixed blocklength and rate.

\begin{example}[A harmless erased ancilla]
\label{ex:physical-guess-versus-logical-recovery}
For the $[[2,1]]_2$ encoder
$|\psi\rangle\mapsto|0\rangle_1\otimes|\psi\rangle_2$, the stabilizer
is generated by $Z_1$.  If $E=\{1\}$, the four physical errors
$I,X_1,Z_1,X_1Z_1$ yield only two syndromes, so the best physical guess
has error $1/2$.  The residual after syndrome correction is a
stabilizer, giving zero logical error.  For $E=\{2\}$ the two errors
are both $3/4$, and for $E=\{1,2\}$ they are $7/8$ and $3/4$,
respectively.  The singleton patterns each have probability $p(1-p)$,
full erasure has probability $p^2$, and no erasure has zero error.
Thus averaging gives
\begin{align}
 \varepsilon_{\mathrm{guess}}^*(\Stab,p)
 &=p(1-p)\left(\frac12+\frac34\right)+p^2\frac78,\\
 \inf_{\cRec\in\mathrm{CPTP}(\mathcal H_B,\mathcal H_A)}
 \varepsilon_{\mathrm{ent}}(\cEnc_{\Stab},\cRec,p)&=p(1-p)\frac34+p^2\frac34,
\end{align}
or, after simplification,
\begin{equation}
 \varepsilon_{\mathrm{guess}}^*(\Stab,p)=\frac{5p}{4}-\frac{3p^2}{8},
 \qquad \inf_{\cRec\in\mathrm{CPTP}(\mathcal H_B,\mathcal H_A)}
 \varepsilon_{\mathrm{ent}}(\cEnc_{\Stab},\cRec,p)=\frac{3p}{4}.
\end{equation}
Indeed, every $[[2,1]]_2$ code has at most two syndromes on a singleton
erasure, so its guessing error is at least $1/2$ there, and is $7/8$
on full erasure.  The code generated by $Z_1Z_2$ attains these bounds:
on either singleton, the restricted syndrome map has rank one, so
$a_{\Stab}(E)+b_{\Stab}(E)=2-1=1$ and the guessing error is $1-2^{-1}$.
On full erasure, $a_{\Stab}([2])+b_{\Stab}([2])=n+k=3$, giving $1-2^{-3}$.
Thus, for $n=2$ and $k=1$,
\begin{equation}\label{eq:optimized-physical-logical-strict-gap}
 \varepsilon_{2,\STAB,\STAB}^*(1/2,p;2)
 \le\frac{3p}{4}<p-\frac{p^2}{8}
 =\varepsilon_{2,\mathrm{guess}}^*(1/2,p;2),\qquad 0<p\le1.
\end{equation}
Increasing the stabilizer dimension can only decrease physical-guessing
error, so the threshold-rate optimum here occurs at $k=1$.
\end{example}

\paragraph{Reliability bounds for physical-error guessing.}
\label{app:guess-prime-q}
We prove the guessing case of
Remark~\ref{prop:common-task-exponent-bounds} for prime $d$,
$0<p<1/2$, and $0<r<1-2p$.  The achievability arguments use the
ensemble $\mu^{\mathrm S}_{n,k}$ from
\eqref{eq:uniform-stabilizer-distribution} and replace nontrivial
logical labels by all nonzero labels in $\Stab^\perp$.

\subparagraph{Random coding.}
For $\Stab\sim\mu^{\mathrm S}_{n,k}$ and a fixed $v\ne0$, the
symplectic-invariance and pair-counting argument in the proof of
Lemma~\ref{lem:random-stabilizer-ambiguity} gives
\begin{equation}\label{eq:guess-label-incidence}
 \pi_{n,k}^{\rm full}:=\Pr\{v\in\Stab^\perp\}
 =\frac{d^{n+k}-1}{d^{2n}-1}\le d^{-(n-k)}.
\end{equation}
The numerator now counts all nonzero labels in $\Stab^\perp$,
including stabilizer labels.  For a fixed erased set $E$, positive
$a_{\Stab}(E)+b_{\Stab}(E)$ means that at least one label in
$V_E\setminus\{0\}$ lies in $\Stab^\perp$.  A union bound yields
\begin{align}
 \mathbb E_{\Stab}\!\left[
   1-d^{-(a_{\Stab}(E)+b_{\Stab}(E))}\right]
 &\le\Pr_{\Stab}\{a_{\Stab}(E)+b_{\Stab}(E)>0\}\notag\\
 &\le(d^{2|E|}-1)\pi_{n,k}^{\rm full}
 \le d^{2|E|-(n-k)}.
\end{align}
The error is also at most one.  Average over $E_p$, choose a code
whose error is no larger than the ensemble average, and set
$k=\lceil nr\rceil$.  Since this code has rate at least $r$,
\begin{equation}\label{eq:guess-finite-random-coding}
 \varepsilon_{n,\mathrm{guess}}^*(r,p;d)
 \le\sum_{e=0}^n\binom ne p^e(1-p)^{n-e}
       \min\{1,d^{2e-(n-k)}\}
 \le d(n+1)e^{-nE_{\rm rc}(r,p;d)}.
\end{equation}
The last inequality is the type estimate in the proof of
Proposition~\ref{thm:random-coding-reliability}.
Taking the lower exponential limit gives
$\underline E_{\mathrm{guess}}(r,p;d)\ge E_{\rm rc}(r,p;d)$.

\subparagraph{Expurgation.}
Replace the logical-label spectrum $L_w(\Stab)$ from
\eqref{eq:logical-weight-enumerator} by
\begin{equation}
 B_w(\Stab):=|\{v\in\Stab^\perp\setminus\{0\}:
 \operatorname{wt}(v)=w\}|,\qquad 1\le w\le n.
\end{equation}
A guessing error can occur only if a nonzero label in
$\Stab^\perp$ is supported on the erased set.  A weight-$w$ label
has this property with probability $p^w$, so the same union bound as
in \eqref{eq:logical-spectrum-union-bound} gives
\begin{equation}
 \varepsilon_{\mathrm{guess}}^*(\Stab,p)
 \le\sum_{w=1}^n B_w(\Stab)p^w.
\end{equation}
There are $\binom nw(d^2-1)^w$ weight-$w$ labels, each with
incidence probability \eqref{eq:guess-label-incidence}.  Hence
\begin{equation}
 \mathbb E_{\Stab} B_w(\Stab)
 =\pi_{n,k}^{\rm full}\binom nw(d^2-1)^w.
\end{equation}
For $1\le w_0\le n$, let
\begin{equation}
 \beta_{n,k,w_0}^{\rm full}
 :=\mathbb E_{\Stab}\!\left[\sum_{w=1}^{w_0-1}B_w(\Stab)\right]
 =\pi_{n,k}^{\rm full}\sum_{w=1}^{w_0-1}\binom nw(d^2-1)^w.
\end{equation}
Apply the conditional-average argument of
Lemma~\ref{prop:finite-logical-expurgation} to
$Z_{w_0}^{\rm full}:=\sum_{w=1}^{w_0-1}B_w(\Stab)$.
Since $Z_{w_0}^{\rm full}$ is a nonnegative integer,
$\Pr\{Z_{w_0}^{\rm full}=0\}\ge1-\beta_{n,k,w_0}^{\rm full}$.
When $\beta_{n,k,w_0}^{\rm full}<1$, nonnegativity gives
\begin{align}
 \mathbb E_{\Stab}\!\left[\sum_{w=w_0}^nB_w(\Stab)p^w
                  \,\middle|\,Z_{w_0}^{\rm full}=0\right]
 &\le\frac{\sum_{w=w_0}^np^w\mathbb E_{\Stab}B_w(\Stab)}
              {\Pr\{Z_{w_0}^{\rm full}=0\}}\notag\\
 &\le\frac{\pi_{n,k}^{\rm full}}{1-\beta_{n,k,w_0}^{\rm full}}
       \sum_{w=w_0}^n\binom nw(d^2-1)^wp^w.
\end{align}
Some code in this event has a spectral sum no larger than its
conditional average.  For that code, $B_w=0$ for $w<w_0$, so
\begin{equation}\label{eq:guess-expurgated-finite-bound}
 \varepsilon_{\mathrm{guess}}^*(\Stab,p)
 \le\frac{\pi_{n,k}^{\rm full}}{1-\beta_{n,k,w_0}^{\rm full}}
 \sum_{w=w_0}^n\binom nw(d^2-1)^wp^w.
\end{equation}
For $k=\lceil nr\rceil$, $w_0=\lceil n\delta\rceil$, and
$0<\delta<\delta_{\rm GV}(r;d)$, the entropy estimate in the proof of
Proposition~\ref{thm:logical-expurgated-reliability} applies because
$\pi_{n,k}^{\rm full}\le d^{-(n-k)}$.  It gives
\begin{equation}
 \beta_{n,k,w_0}^{\rm full}\le dn
 e^{-n((1-r)\ln d-H_{d^2}(\delta))}\longrightarrow0.
\end{equation}
Bounding each summand in \eqref{eq:guess-expurgated-finite-bound}
by its type exponent and summing at most $n$ terms gives
\begin{equation}
 \varepsilon_{n,\mathrm{guess}}^*(r,p;d)
 \le\min\left\{1,\frac{dn}{1-\beta_{n,k,w_0}^{\rm full}}
 e^{-n[(1-r)\ln d-\max_{\delta\le\omega\le1}
       \{H_{d^2}(\omega)+\omega\ln p\}]}\right\}.
\end{equation}
The prefactor has vanishing normalized logarithm.  Taking the lower
exponential limit and then $\delta\uparrow\delta_{\rm GV}(r;d)$
therefore yields
$\underline E_{\mathrm{guess}}(r,p;d)\ge E_{\rm expur}(r,p;d)$.

\subparagraph{Converse and matching.}
By \eqref{eq:guess-logical-ordering}, optimal guessing error is at
least optimal logical recovery error.  Taking normalized negative
logarithms and then upper limits, using the class ordering in
\eqref{eq:code-class-exponent-ordering}, and applying
Proposition~\ref{prop:reliability-converse}, gives, with arguments $(r,p;d)$
suppressed,
\begin{equation}
 \overline E_{\mathrm{guess}}
 \le\overline E_{\STAB,\STAB}
 \le\overline E_{\All,\All}
 \le\Dbin\!\left(\frac{1-r}{2}\middle\|p\right).
\end{equation}
Together with the two lower bounds, this proves the guessing case of
Remark~\ref{prop:common-task-exponent-bounds}.
When $r\ge r_{\rm crit}(p,d)$,
Proposition~\ref{prop:random-coding-closed-form} identifies
$E_{\rm rc}(r,p;d)$ with this upper bound.  The lower and upper
exponents are then equal, so $E_{\mathrm{guess}}(r,p;d)$ exists and equals
$\Dbin((1-r)/2\|p)$.

The full and logical incidence probabilities have the same exponential
order at positive rate.  Indeed, by
\eqref{eq:logical-label-incidence-probability} and
\eqref{eq:guess-label-incidence},
\begin{equation}
 \frac{\pi_{n,k}^{\rm full}}{\pi_{n,k}}
 =\frac{1-d^{-(n+k)}}{1-d^{-2k}}\longrightarrow1
 \quad\text{when }k=\lceil nr\rceil,\ r>0.
\end{equation}
Thus the same expurgated lower bound holds for the guessing error exponent
and all four quantum reliability functions.  These bounds do not determine
whether the guessing exponent equals each quantum reliability function or
is strictly smaller below the critical
rate, even though individual-code and finite-blocklength errors can differ.

\subsection{No-feedback classical communication}
\paragraph{Task definition and error exponents.}

\label{subsec:classical-nf}
Consider uniformly distributed classical messages sent over the
auxiliary $d^2$-dimensional erasure channel without feedback or
preshared entanglement.  Following
\cite[Section~2.2, Eq.~(2.5)]{wildeFiniteblocklengthClassicalCommunication2026},
the encoder assigns density operators $\rho_j$ on
$(\mathbb C^{d^2})^{\otimes n}$, possibly entangled across inputs,
and the receiver uses a joint positive operator-valued measure
(POVM) $\{\Lambda_j\}_{j=1}^{M_n}$ on the flagged output.
This means $\Lambda_j\ge0$, $\sum_j\Lambda_j=I$, and outcome $j$
is the decoded message; on output state $\tau$, it has probability
$\tr(\Lambda_j\tau)$.  The measurement acts jointly on all $n$ outputs.  Put $k_n=\lceil nr\rceil$ and $M_n=d^{n+k_n}$, the number
of equiprobable messages, and define
\begin{equation}\label{eq:classical-nf-finite-error}
\begin{aligned}
 \varepsilon_{n,\mathrm{cl}}^*(r,p;d)
 &:=\varepsilon_{\mathrm{nf}}^*(n,M_n;d^2,p)\\
 &=\min_{\{\rho_j\},\{\Lambda_j\}}
 \left[1-\frac1{M_n}\sum_{j=1}^{M_n}
 \tr\!\left(\Lambda_j\cEra_{p,d^2}^{\otimes n}(\rho_j)\right)\right].
\end{aligned}
\end{equation}
The index $\mathrm{cl}$ denotes this no-feedback task, which allows
quantum code states.  Its asymptotic message rate is
$R_{\mathrm{cl}}=(1+r)\ln d$ nats per channel use.
Wilde's no-feedback reliability bounds
\cite[Proposition~8.7]{wildeFiniteblocklengthClassicalCommunication2026}
apply with $d_{\rm W}=d^2$ and classical rate $(1+r)\log_2d$ bits per channel
use.  The calculations below detail the conversion of
the exponent bounds and critical rate and check the message-count
convention.

For this task, define the lower and upper error exponents by
\begin{equation}\label{eq:classical-nf-reliability-exponents}
\begin{aligned}
 \underline E_{\mathrm{cl}}(r,p;d)
 &:=\liminf_{n\to\infty}-\frac1n\ln\varepsilon_{n,\mathrm{cl}}^*(r,p;d),\\
 \overline E_{\mathrm{cl}}(r,p;d)
 &:=\limsup_{n\to\infty}-\frac1n\ln\varepsilon_{n,\mathrm{cl}}^*(r,p;d).
\end{aligned}
\end{equation}
We use $-\ln0=+\infty$ and denote the common value by
$E_{\mathrm{cl}}(r,p;d)$ only when these limits coincide.

\paragraph{Parameter conversion.}\label{app:classical-nf-exponent}

We verify how Wilde's Proposition~8.7
\cite{wildeFiniteblocklengthClassicalCommunication2026} yields the
classical-message case of Remark~\ref{prop:common-task-exponent-bounds}
and Remark~\ref{cor:degeneracy-high-rate}.  Let $d$ be prime,
$0<p<1/2$, and $0<r<1-2p$.  The model in
\eqref{eq:classical-nf-finite-error} allows arbitrary input states
and a joint decoding POVM, without preshared entanglement, as in
\cite[Section~2.2]{wildeFiniteblocklengthClassicalCommunication2026}.
Denote Wilde's local dimension by $d_{\rm W}$; in his notation, take
\begin{equation}\label{eq:classical-nf-wilde-parameters}
 d_{\rm W}=d^2,\qquad R_{\rm W}=(1+r)\log_2d,\qquad
 \frac{R_{\rm W}}{\log_2d_{\rm W}}=\frac{1+r}{2}.
\end{equation}
Here $R_{\rm W}$ is the classical rate in bits per channel use, so
$0<R_{\rm W}<(1-p)\log_2d_{\rm W}$.  Multiplying Wilde's exponents by
$\ln2$ converts them to nats.\footnote{Our message count
$M_n=d^{n+\lceil nr\rceil}$ satisfies
$n^{-1}\log_2M_n=R_{\rm W}+O(n^{-1})$.
In the cited proof, this rate perturbation changes the random-coding
objective by $O(n^{-1})$, and the expurgated exponent by $O(n^{-1})$
for each fixed expurgation parameter.  The normalized binomial-tail
threshold tends to $(1+r)/2$.  Thus the same bounds hold for our
sequence by taking the corresponding limits, and then the supremum
over fixed expurgation parameters; see
\cite[Eqs.~(8.37)--(8.38), (8.45), and (B.4)]{wildeFiniteblocklengthClassicalCommunication2026}.}

\paragraph{Random-coding bound.}
Let $E_{\rm r}$ denote Wilde's random-coding exponent
\cite[Eq.~(8.40)]{wildeFiniteblocklengthClassicalCommunication2026}.
Writing his survivor fraction as $u=1-t$, the symmetry
$\Dbin(1-t\|1-p)=\Dbin(t\|p)$ and
$R_{\rm W}\ln2=(1+r)\ln d$ give
\begin{equation}
\begin{aligned}
 (\ln2)E_{\rm r}(R_{\rm W})
 &=\min_{0\le u\le1}\left\{
   \Dbin(u\|1-p)+(u\ln d_{\rm W}-R_{\rm W}\ln2)_+\right\}\\
 &=\min_{0\le t\le1}\left\{
   \Dbin(t\|p)+\big(2(1-t)-(1+r)\big)_+\ln d\right\}\\
 &=E_{\rm rc}(r,p;d),
\end{aligned}
\end{equation}
where the last equality is \eqref{eq:random-coding-exponent}.

\paragraph{Expurgated bound.}
The spectral expression for $E_{\rm expur}$ can also be written as an
optimization over a single auxiliary parameter.  This equivalent
Gallager form is useful for comparison with expurgated bounds for other
communication tasks; see Gallager's low-rate bound for the classical
method \cite[Section~V]{gallagerSimpleDerivation1965}.

\begin{prop}[Gallager form of the expurgated bound]
\label{prop:logical-expurgated-gallager}
For $0<p<1/2$ and $0\le r<1-2p$, the expurgated exponent in
\eqref{eq:logical-expurgated-spectral} has the equivalent form
\begin{equation}\label{eq:logical-expurgated-gallager}
 E_{\rm expur}(r,p;d)
 =\left[\sup_{\rho\ge1}\rho\left\{(1-r)\ln d
 -\ln\!\left(1+(d^2-1)p^{1/\rho}\right)\right\}\right]_+.
\end{equation}
\end{prop}
\begin{proof}
For $0<\omega<1$,
\begin{equation}
 \frac{d}{d\omega}[H_{d^2}(\omega)+\omega\ln p]
 =\ln\frac{(d^2-1)p(1-\omega)}{\omega}.
\end{equation}
This derivative is positive below and negative above $\omega_p$ in
\eqref{eq:expurgated-omega-p}.
Abbreviate $\delta_{\rm GV}:=\delta_{\rm GV}(r;d)$.
If $\delta_{\rm GV}\ge\omega_p$, the constrained maximum is attained
at $\delta_{\rm GV}$, giving
\begin{equation}
 E_{\rm expur}(r,p;d)
 =[(1-r)\ln d-H_{d^2}(\delta_{\rm GV})-\delta_{\rm GV}\ln p]_+
 =\delta_{\rm GV}\ln(1/p)
\end{equation}
by \eqref{eq:logical-gv-distance}.  Otherwise it is attained at
$\omega_p$, where
$H_{d^2}(\omega_p)+\omega_p\ln p=\ln(1+(d^2-1)p)$.
For the function inside the supremum in
\eqref{eq:logical-expurgated-gallager}, write
\begin{equation}
 f(\rho):=\rho\bigl[(1-r)\ln d
             -\ln(1+(d^2-1)p^{1/\rho})\bigr].
\end{equation}
Put
\begin{equation}
 \theta_\rho:=\frac{(d^2-1)p^{1/\rho}}{1+(d^2-1)p^{1/\rho}}.
\end{equation}
The identity
\begin{equation}
 H_{d^2}(\theta_\rho)
 =\ln(1+(d^2-1)p^{1/\rho})
   -\frac{\theta_\rho}{\rho}\ln p
\end{equation}
then gives
\begin{equation}
 f'(\rho)=(1-r)\ln d-H_{d^2}(\theta_\rho),\qquad
 \theta_\rho'=-\frac{\ln p}{\rho^2}\theta_\rho(1-\theta_\rho)>0.
\end{equation}
As $\rho$ increases from $1$ to infinity, $\theta_\rho$ increases from
$\omega_p$ to $1-d^{-2}$.  Entropy is strictly increasing on this
interval, so $f'$ is strictly decreasing, with negative limiting
value $-(1+r)\ln d$.
If $\delta_{\rm GV}\ge\omega_p$, the derivative changes sign when
$\theta_\rho=\delta_{\rm GV}$; equality at the initial point means
the maximum occurs at $\rho=1$.  Substituting the preceding entropy
identity and $H_{d^2}(\delta_{\rm GV})=(1-r)\ln d$ gives, at that maximum,
\begin{equation}
 f(\rho)=\rho[(1-r)\ln d-H_{d^2}(\delta_{\rm GV})]
           -\delta_{\rm GV}\ln p
 =\delta_{\rm GV}\ln(1/p).
\end{equation}
If $\delta_{\rm GV}<\omega_p$, then $f'(1)<0$, and $f$ decreases
throughout $[1,\infty)$.  Its maximum is
$f(1)=(1-r)\ln d-\ln(1+(d^2-1)p)$.  Taking positive parts gives
\eqref{eq:logical-expurgated-gallager} in both cases.
\end{proof}

Let $E_{\rm ex}$ denote Wilde's expurgated exponent
\cite[Eq.~(8.41)]{wildeFiniteblocklengthClassicalCommunication2026}.
For every fixed $\rho\ge1$, substitution of $d_{\rm W}=d^2$ gives
\begin{equation}
 \frac1{d_{\rm W}}+\left(1-\frac1{d_{\rm W}}\right)p^{1/\rho}
 =d^{-2}\left(1+(d^2-1)p^{1/\rho}\right).
\end{equation}
Consequently,
\begin{equation}
\begin{aligned}
 (\ln2)E_{\rm ex}(R_{\rm W})
 &=\sup_{\rho\ge1}\left\{
 -\rho\ln\!\left[\frac1{d_{\rm W}}+\left(1-\frac1{d_{\rm W}}\right)p^{1/\rho}\right]
 -\rho R_{\rm W}\ln2\right\}\\
 &=\sup_{\rho\ge1}\rho\left\{
  2\ln d-\ln\!\left(1+(d^2-1)p^{1/\rho}\right)-(1+r)\ln d\right\}\\
 &=\sup_{\rho\ge1}\rho\left\{
  (1-r)\ln d-\ln\!\left(1+(d^2-1)p^{1/\rho}\right)\right\}.
\end{aligned}
\end{equation}
By \eqref{eq:logical-expurgated-gallager},
$E_{\rm expur}(r,p;d)=[(\ln2)E_{\rm ex}(R_{\rm W})]_+$.
An optimal error exponent is nonnegative, so the positive part
preserves the lower bound.

\paragraph{Converse and high-rate condition.}
The converse in
\cite[Eq.~(8.43)]{wildeFiniteblocklengthClassicalCommunication2026}
becomes, in nats,
\begin{equation}
 \Dbin\!\left(\frac{R_{\rm W}}{\log_2d_{\rm W}}\middle\|1-p\right)
 =\Dbin\!\left(\frac{1+r}{2}\middle\|1-p\right)
 =\Dbin\!\left(\frac{1-r}{2}\middle\|p\right).
\end{equation}
Together with the preceding two calculations, this gives exactly
the classical-message bounds in
\eqref{eq:ostrev-errorguess-exponent-sandwich}.
The critical-rate condition in
\cite[Eqs.~(8.39) and (8.44)]{wildeFiniteblocklengthClassicalCommunication2026}
transforms as
\begin{equation}
\begin{aligned}
 R_{\rm W}\ge\frac{1-p}{1-p+d_{\rm W}p}\log_2d_{\rm W}
 &\quad\Longleftrightarrow\quad
 \frac{1+r}{2}\ge\frac{1-p}{1-p+d^2p}\\
 &\quad\Longleftrightarrow\quad
 r\ge\frac{1-(d^2+1)p}{1+(d^2-1)p}=r_{\rm crit}(p,d).
\end{aligned}
\end{equation}
Thus Wilde's high-rate equality supplies the classical-message
exponent in Remark~\ref{cor:degeneracy-high-rate}.
The guessing error exponent and quantum reliability functions follow from
the reliability-bounds paragraph in
Appendix~\ref{app:guess-prime-q},
Proposition~\ref{prop:random-coding-closed-form}, and
Corollary~\ref{prop:reliability-matching}.
\subsection{Common reliability bounds and equality above the critical rate}
The task index $\gamma$ ranges over $\mathrm{cl}$, $\mathrm{guess}$,
and the four quantum classes $(\STAB,\STAB)$, $(\STAB,\All)$,
$(\All,\STAB)$, and $(\All,\All)$.
\begin{remark}[Common bounds for the coding tasks]
\label{prop:common-task-exponent-bounds}
\label{prop:ostrev-errorguess-exponent}
For prime $d$, $0<p<1/2$, and $0<r<1-2p$, every task $\gamma$ satisfies
\begin{equation}\label{eq:ostrev-errorguess-exponent-sandwich}
 \max\{E_{\rm rc}(r,p;d),E_{\rm expur}(r,p;d)\}
 \le\underline E_{\gamma}(r,p;d)
 \le\overline E_{\gamma}(r,p;d)
 \le\Dbin\!\left(\frac{1-r}{2}\middle\|p\right).
\end{equation}

The bounds for all four quantum classes follow from
Theorem~\ref{thm:reliability-main} and the class ordering in
\eqref{eq:code-class-exponent-ordering}.
The reliability-bounds paragraph in Appendix~\ref{app:guess-prime-q}
proves both guessing lower bounds;
\eqref{eq:guess-logical-ordering} and
Proposition~\ref{prop:reliability-converse} give its upper bound.
For $\gamma=\mathrm{cl}$, Wilde's Proposition~8.7
\cite{wildeFiniteblocklengthClassicalCommunication2026} gives both
lower bounds and the upper bound after the parameter conversion in the
corresponding paragraph of Appendix~\ref{app:classical-nf-exponent}.
For every task, the liminf is at most the limsup.
\end{remark}

\begin{remark}[Equality above the critical rate]
\label{cor:degeneracy-high-rate}
For prime $d$, $0<p<1/2$, and $0<r<1-2p$ with
$r\ge r_{\rm crit}(p,d)$, all four quantum reliability functions and
the guessing and classical error exponents exist, and for every task $\gamma$,
\begin{equation}\label{eq:guess-coset-high-rate}
 E_{\gamma}(r,p;d)=\Dbin\!\left(\frac{1-r}{2}\middle\|p\right).
\end{equation}

In this rate range, Proposition~\ref{prop:random-coding-closed-form}
gives $E_{\rm rc}(r,p;d)=\Dbin((1-r)/2\|p)$.
Remark~\ref{prop:common-task-exponent-bounds} therefore gives,
for each task,
\begin{equation}
 \Dbin((1-r)/2\|p)\le\underline E_{\gamma}(r,p;d)
 \le\overline E_{\gamma}(r,p;d)\le\Dbin((1-r)/2\|p).
\end{equation}
The lower and upper limits coincide, proving the claim.
\end{remark}

The common value reflects the same erasure threshold
$t=(1-r)/2$.  At or above the critical rate, the random-coding minimum
occurs at this threshold, where its collision penalty vanishes.
The quantum change-of-measure converse and the mapped classical
dimension bound approach the same threshold; its binomial
large-deviation cost is $\Dbin((1-r)/2\|p)$.
Thus the physical-error guessing exponent already equals the quantum
reliability function in this region.  The equality concerns
optimized asymptotic quantities at the specified parameters; finite-blocklength
errors and individual-code performance can differ, as illustrated in
Appendix~\ref{app:ostrev}.  Below the critical rate, the common bounds
alone do not determine whether the reliability functions and error exponents coincide.

\end{document}